\documentclass[twocolumn,showpacs,preprintnumbers,amsmath,amssymb,psamsfonts]{revtex4}
\usepackage{color}
\usepackage{amscd}
\usepackage{xspace}
\newtheorem{corollary}{Corollary}
\newtheorem{lemma}{Lemma}
\newtheorem{example}{Example}

\renewcommand{\Re}{\operatorname{Re}}
\renewcommand{\Im}{\operatorname{Im}}

\newcommand{\N}{\ensuremath{\mathbb N}{}}
\newcommand{\R}[1]{\ensuremath{\mathbb R}^{\,#1}{}}
\newcommand{\C}[1]{\ensuremath{\mathbb C}^{\,#1}{}}
\newcommand{\Z}{\mathbb{Z}}
\newcommand{\unity}{\ensuremath{{\rm 1 \negthickspace l}{}}}

\newcommand{\ket}[1]{\ensuremath{| #1 \rangle}{}}

\newcommand{\maxn}{\operatorname{max}}

\newcommand{\Mat}{\operatorname{Mat}{}}
\newcommand{\diag}{\operatorname{diag}{}}

\newcommand{\minover}[1]{\ensuremath{\underset{#1}\min}}

\newcommand{\grape}{{\sc grape}\xspace}
\newcommand{\risc}{\text{\sc risc}\xspace}
\newcommand{\cisc}{\text{\sc cisc}\xspace}
\newcommand{\cpu}{{\sc cpu}\xspace}
\newcommand{\qnot}{{\sc not}\xspace}
\newcommand{\gnot}{{\sc not}\xspace}
\newcommand{\cnot}{{\sc cnot}\xspace}
\newcommand{\cnots}{{\sc cnot}s\xspace}
\newcommand{\cnnot}{{\sc c}\ensuremath{^n}{\sc not}\xspace}
\newcommand{\ctnot}{{\sc c}\ensuremath{^2}{\sc not}\xspace}
\newcommand{\cmtnot}{{\sc c}\ensuremath{^{n-2}}{\sc not}\xspace}
\newcommand{\cjnot}[1]{\text{\sc c}\ensuremath{^{#1}}\text{\sc not}\xspace}
\newcommand{\cjnotk}[2]{\text{\sc c}\ensuremath{^{#1}}\text{\sc not}\ensuremath{^{#2}}\xspace}
\newcommand{\cjnots}[1]{\text{\sc c}\ensuremath{^{#1}}\text{\sc not}s\xspace}
\newcommand{\cttnot}{{\sc c}\ensuremath{^3}{\sc not}\xspace}

\newcommand{\cmnot}{{\sc c}\ensuremath{^m}{\sc not}\xspace}

\newcommand{\cjU}[1]{\text{\sc c}\ensuremath{^{#1}}\text{\sc U}\xspace}
\newcommand{\qft}{\text{\sc qft}\xspace}
\newcommand{\dft}{\text{\sc dft}\xspace}
\newcommand{\qfts}{\text{\sc qft}s\xspace}

\newcommand{\qftm}{\text{{\sc qft}}\ensuremath{_m}\xspace}
\newcommand{\swap}{\text{\sc swap}\xspace}
\newcommand{\swapj}[1]{\text{{\sc swap}}\ensuremath{_{1,#1}}\xspace}
\newcommand{\swapsj}[1]{\text{{\sc swap}s}\ensuremath{_{1,#1}}\xspace}
\newcommand{\swaps}{\text{\sc swap}s\xspace}
\newcommand{\cpswap}{\text{c{\sc p-swap}}\xspace}
\newcommand{\cpswaps}{\text{c{\sc p-swap}}s\xspace}
\newcommand{\cpswapjm}{\text{c{\sc p-swap}}\ensuremath{_m^j}\xspace}

\newcommand{\cpswapJM}[2]{\text{c{\sc p-swap}}\ensuremath{_{#2}^{#1}}\xspace}

\newcommand{\hlrbii}{\text{\sc hlrb-ii}\xspace}

\renewcommand{\mod}{\operatorname{mod}{\;}}

\newcommand{\fnormsq}[1]{\ensuremath{\Vert #1 \Vert{}}_2^2}

\newcommand{\floor}[1]{\ensuremath{\lfloor #1 \rfloor{}}}
\newcommand{\ceil}[1]{\ensuremath{\lceil #1 \rceil{}}}

\newcommand{\tr}{\operatorname{tr}}

\usepackage{epic}
\usepackage{german}
\usepackage{type1cm}

\usepackage{graphicx}
\usepackage{dcolumn}
\usepackage{bm}

\begin{document}


\title{Quantum CISC Compilation by Optimal Control and \\[1mm]
Scalable Assembly of Complex Instruction Sets beyond Two-Qubit Gates}

\author{T.~Schulte-Herbr{\"u}ggen}\email{tosh@ch.tum.de}
\author{A.~Sp{\"o}rl}
\author{S.J.~Glaser}
\affiliation{Department of Chemistry, %
	Technical University Munich, D-85747 Garching, Germany}

\date{\today}


\begin{abstract}
We present a quantum \cisc compiler and show how to assemble complex instruction sets
in a scalable way. Enlarging the toolbox of universal gates by optimised complex multi-qubit
instruction sets thus paves the way to fight relaxation for realistic experimental settings.

Compiling a quantum module into the machine code for steering a concrete quantum hardware
device lends itself to be tackled by means of optimal quantum control. 
To this end, there are two opposite approaches: (i) one may use
a decomposition into the restricted instruction set (\risc) of 
universal one- and two-qubit gates and translate them into the machine code,
or (ii) one may prefer to generate the entire target module
as a complex instruction set (\cisc) 
directly by evoltution under drift and available controls. 
Here we advocate direct compilation up to the limit of system size
a classical high-performance parallel computer cluster can reasonably handle. 
For going beyond these limits, i.e. for large systems, we propose a combined way,
namely (iii) to make recursive use of medium-sized building blocks generated by optimal control
in the sense of a quantum \cisc compiler. 

The advantage of the method over standard \risc compilations into one- and two-qubit
universal gates is explored on the parallel cluster \hlrbii (with a total {\sc linpack}
performance of $63.3$ TFlops/s) for the quantum Fourier transform, the indirect \swap gate
as well as for multiply-controlled \gnot gates.
Implications for upper limits to time complexities are also derived.

\end{abstract}

\maketitle

\section*{Introduction}
Richard Feynman's seminal conjecture of using experimentally controllable quantum systems to
perform computational tasks \cite{Fey82, Fey96} roots in reducing the complexity of 
the problem when moving from a classical setting to a quantum setting. The most prominent
pioneering example being Shor's quantum algorithm of prime factorisation \cite{Shor94, Shor97}
which is of polynomial complexity ({\sc bqp}) on quantum devices instead of showing
non-polynomial complexity on classical ones \cite{Pap95}.
It is an example of a class of quantum algorithms \cite{Jozsa88, Mosca88} that solve
{\em hidden subgroup problems} in an efficient way \cite{EHK04}, where
in the Abelian case, the speed-up hinges on the quantum Fourier transform ({\qft}).
Whereas the network complexity of the fast Fourier transform ({\sc fft}) for $n$ classical bits
is of order $O(n 2^n)$ \cite{CT65, Beth84}, the {\sc qft} for $n$ qubits shows a complexity
of order $O(n^2)$.
Moreover, Feynman's second observation that quantum systems may be used to efficiently 
predict the behaviour of other quantum systems has inaugurated
a branch of research dedicated to Hamiltonian 
simulation \cite{Lloyd96,AL97,Zal98,BCL+02,MVL02,JC03}.

For implementing a quantum algorithm in an experimental setup, local operations and
universal two-qubit quantum gates are required as a minimal set ensuring every unitary
module can be realised \cite{Deu85}. 
More recently, it turned out that generic qubit and qudit pair interaction
Hamiltonians suffice to complement local actions to universal controls \cite{DNB+02,BBN05}.
Common sets of quantum computational instructions
comprise (i) local operations such as the Hadamard gate, the phase gate and (ii) the
entangling operations {\sc cnot}, controlled-phase gates,
$\sqrt{\text{\sc swap}}$, $i$\,{\sc swap} as well as (iii) the {\sc swap} operation.
The number of elementary gates required for implementing
a quantum module then gives the network or gate complexity.

As is well known, a generic $n$-qubit generic operation requires exponentially many 
two-qubit gates to be implemented exactly \cite{Barenco, Knill95a}, the complexity 
being $O(n 4^n)$. Yet, as has been
pointed out by Barenco {\em et al.}, many quantum computationally pertinent gates
can be decomposed into a number of one- and two-qubit gates {\em increasing linearly} with the number
of qubits. At the expense of a single ancilla qubit this also holds for
multiply controlled unitary gates \cite{Barenco} tantamount to error correction.
For an overview, see e.g.~\cite{NC00,Kit02,Mer07}. 
Moreover, Blais \cite{Blais} showed how to implement the QFT with linear gate
complexity. Later, Solovay (\cite{Solovay95} quoted in \cite{Kit97} and \cite{NC00}) 
and then Kitaev addressed the problem to approximate
arbitrary unitary gates by polynomially long 2-qubit gate sequences 
up to a given precision \cite{Kit97, Kit02}.
More recently the bounds of approximating an arbitrary unitary
were taken down to a polynomial of sixth-order in the number of qubits
and of third order in the geodesic distance of the unitray to unity \cite{NDG+06}.
Differential geometric aspects in terms of Finsler metrics have been raised in \cite{N06}.
\begin{figure*}[Ht!]
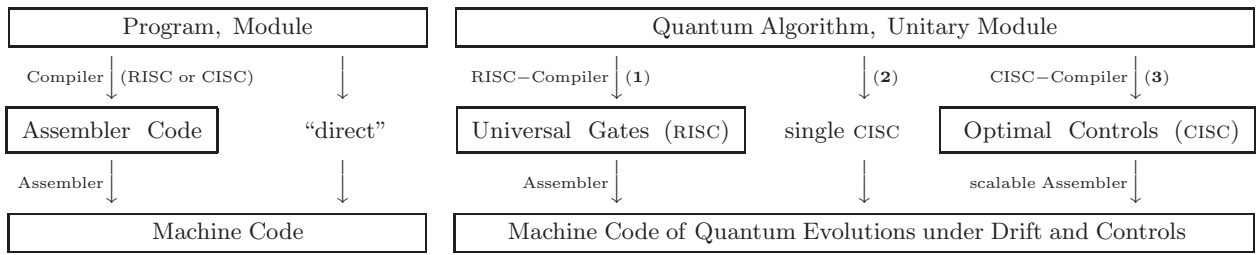

\begin{center}
\hspace{-12mm}\boxed{\rm \hspace{14mm} Program,\; Module \hspace{14mm}}
\hspace{1.5mm}
\boxed{\rm \hspace{25mm} Quantum\; Algorithm,\; Unitary\; Module \hspace{25mm}}
\end{center}
\vspace{-11mm}
\begin{center}
\begin{equation*}
\begin{CD}
\hspace{-25mm} @V {\rm \hspace{-8mm} Compiler} V {\rm (RISC\; or \;CISC)} V \hspace{-37mm} @VVV \hspace{-60mm} @V {\rm {RISC}-Compiler} V{\rm \bf (1)}V %
		\hspace{-62mm} @VV{\rm \bf (2)}V \hspace{-17mm} @V{\rm {CISC}-Compiler}V{\rm \bf (3)}V \hspace{2mm}\\
\hspace{-25mm} \boxed{\;\text{Assembler\phantom{$(^!$}Code}\;} &&  %
	\hspace{8mm}\text{\quad \/``direct\/''} \hspace{9mm}
        \boxed{\;\text{Universal\phantom{$(^!$}Gates\;\;({\sc risc})}\;} &&  %
	\text{\quad\; {single {\sc cisc} \quad\;}} && %
	\boxed{\text{\; {Optimal\phantom{$(^!$}Controls\; ({\sc cisc})\;}}}\\
\hspace{-25mm} @V{\rm Assembler}VV \hspace{-37mm} @VVV \hspace{-60mm} @V{\rm Assembler}VV \hspace{-62mm} %
		@VVV \hspace{-17mm} @V{\rm scalable\; Assembler}VV \hspace{2mm}\\
\end{CD}
\vspace{-2mm}
\end{equation*}
\hspace{-12mm}\boxed{\rm \hspace{14.2mm} \phantom{,,,} Machine\;Code\; \hspace{14.2mm}}
\hspace{1.5mm}
\boxed{\rm \hspace{6mm} Machine\;Code\; of\; Quantum\; Evolutions\; under\; Drift\; and \; Controls
\phantom{,}\hspace{7mm}}
\caption{\label{fig:CISQC-compile}
Compilation in classical computation (left) and quantum computation (right).
Quantum machine code has to be time-optimal or 
protected against relaxation, otherwise the coherent 
superpositions are wiped out. 
A  quantum \risc-compiler (1) by universal gates
leads to unnecessarily long machine code. Direct \cisc-compilation into a single
pulse sequence (2) exploits quantum control for a near time-optimal quantum machine code.
Its classical complexity is {\sc np}, so direct compilation by numerical optimal control
resorting to a classical computer is unfeasible for large quantum systems.
The third way (3) promoted here pushes quantum \cisc-compilation
to the limits of classical supercomputer clusters and then assembles the multi-qubit complex instructions sets recursively
into time-optimised or relaxation-protected quantum machine code.
}
\end{center}
\end{figure*}

However, gate complexity often translates into too coarse an estimate for the actual
time required to implement a quantum module (see e.g. \cite{VHC02, CHN03, ZGB}),
in particular, if the time scales of a specific experimental setting have to be matched.
Instead, effort has been taken to give upper bounds on the actual {time complexity} \cite{WJB02},
e.g., by way of numerical optimal control \cite{PRA05}.

Interestingly, in terms of 
quantum control theory, the {\em existence of universal gates} is equivalent to the
statement that the quantum system is {\em fully controllable} as has first
been pointed out in Ref.~\cite{RaRa95}. This is, e.g., the case in systems of
$n$ spin-$\tfrac{1}{2}$ qubits that form Ising-type weak-coupling topologies described by arbitrary
connected graphs \cite{TOSH-Diss,Science98}.
Therefore the usual approach to quantum compilation in terms of local plus universal two-qubit
operations \cite{Tucci99,Will04,SBM06,Chuang06,Tucci07} lends itself to be complemented by optimal-control
based direct compilation into machine code: it may be seen as a technology-dependent optimiser in the sense
of Ref.~\cite{Chuang06}, however, tailored to deal with more complex instruction sets than the usual
local plus two-qubit building blocks. 
Not only is it adapted to the specific experimental setting, it also allows for fighting relaxation
by either being near timeoptimal or by exploiting relaxation-protected subspaces \cite{PRL_decoh}.
Devising quantum compilation methods for optimised
realisations of given quantum algorithms by admissible controls is therefore an
issue of considerable practical interest. 
Here it is the goal to show how quantum compilation can favourably be accomplished
by optimal control: the building blocks for gate synthesis will be extended from the usual set
of restricted local plus universal two-qubit gates to a larger toolbox of {\em scalable} multi-qubit
gates tailored to yield high fidelity in short time given concrete experimental settings.

\section*{Quantum Compilation as an Optimal Control Task}
As shown in Fig.~\ref{fig:CISQC-compile},
the quantum compilation task can be addressed following different principle guidelines:
{\bf (1)} by the standard decomposition into local operations and universal two-qubit gates,
which by analogy to classical computation was termed {\em reduced instruction set} quantum
computation (\risc) \cite{SKH98} or
{\bf (2)} by using direct compilation into one single 
{\em complex instruction set} (\cisc) \cite{SKH98}. 
The existence of a such a single effective gate is guaranteed simply by the
unitaries forming a group: a sequence of local plus universal gates is a product
of unitaries and thus a single unitary itself. 

As a consequence, \cisc quantum compilation lends itself
for resorting to numerical optimal control (on clusters of classical computers)
for translating the unitary target module directly into the
\/`machine code\/' of evolutions of the quantum system under combinations of the drift
Hamiltonian $H_0$ and experimentally available controls $H_j$. 

In a number of studies on
quantum systems up to 10 qubits, we have shown that direct compilation by gradient-assisted 
optimal control \cite{GRAPE,PRA05,PRA07} allows for substantial speed-ups, e.g., by a factor
of $5$ for a \cnot and a factor of $13$ for a Toffoli-gate on coupled Josephson qubits \cite{PRA07}.
However, the direct approach naturally faces the limits of computing quantum systems on classical devices: 
upon parallelising our {C++} code for high-performance clusters \cite{EP06},
we found that extending the quantum system by one qubit increases the \cpu-time required for 
direct compilation into the quantum machine code of controls by roughly a factor of eight.
So the classical complexity for optimal-control based quantum compilation is {\sc np}.

Therefore, here we advocate a third approach {\bf (3)} that uses direct compilation into
multi-qubit complex instruction sets up
to the \cpu-time limits of optimal quantum control on classical computers: 
these building blocks are designed such as to allow for recursive scalable quantum compilation
in large quantum systems (i.e. those beyond classical computability). In particular, the
complex instruction sets may be optimised such as to fight relaxation by being near time-optimal,
or, moreover, they may be devised such as to fight the specific imperfections of an experimental
setting.

\subsubsection*{Controllability}
Before turning to optimal-control based \cisc quantum compilation in more detail, it is important to ensure
the quantum control system characterised by $\{H_0\} \cup \{H_j\}$ is in fact
{\em fully controllable}.

Hamiltonian quantum dynamics following Schr{\"o}dinger's equation
for the unitary image of a complete basis set of \/`state vectors\/'
representing a quantum gate
\begin{eqnarray}
        \ket{\dot\psi(t)} &=& -i\big(H_d + \sum_{j=1}^m u_j(t) H_j\big) \;\ket{\psi(t)}\\
        \label{eqn:bilinear_contr}
        {\dot U(t)} &=& -i\big(H_d + \sum_{j=1}^m u_j(t) H_j\big) \;{U(t)} \quad,
\end{eqnarray}
resembles the setting of a standard {\em bilinear control system} with state $X(t)$,
drift $A$, controls $B_j$, and control amplitudes $u_j\in\R{}$
reading
\begin{equation}
        \dot X(t) = \big(A + \sum_{j=1}^m u_j(t) B_j\big) \; X(t) \quad,
\end{equation}
where $X(t) \in {GL}_N(\C{})$ and $A,B_j\in \Mat_N(\C{})$.
Clearly in the dynamics of closed quantum systems, the system Hamiltonian $H_d$ is the drift term,
whereas the $H_j$ are the control Hamiltonians with $u_j(t)$ as control amplitudes.
In systems of $n$ qubits, $\ket{\psi}\in \C{2^n}$,
$U\in SU(2^n)$, and $i\,H_\nu\in\mathfrak{su}(2^n)$.

A system is {\em fully operator controllable}, if to every initial state $\rho_0$ the entire unitary
orbit $\mathcal O_{\rm U}(\rho_0):=\{U\rho_0 U^\dagger\;|\; U\in SU(N)\}$ can be reached.
With density operators being Hermitian
this means any final state $\rho(t)$ can be reached from any initial state $\rho_0$
as long as both of them share the same spectrum of eigenvalues. 

As established in \cite{JS72},
the bilinear system of Eqn.~\ref{eqn:bilinear_contr} is {fully controllable} if and only if
the drift and controls are a generating set of $\mathfrak{su}(N)$ by
way of the commutator, i.e.,
$\langle{H_d, H_j} \,|\,j=1,2,\dots,m\rangle_{\rm Lie} = {\mathfrak{su}(N)}$.

\medskip
{\bf Example 1}
Consider a system of $n$ weakly coupled spin-$\tfrac{1}{2}$ qubits.
Let $\sigma_x = \left(\begin{smallmatrix} \,0 &\,1\, \\ \,1 &\,0\, \end{smallmatrix}\right)$,
$\sigma_y = \left(\begin{smallmatrix} 0 &-i \\ i &\phantom{-}0 \end{smallmatrix}\right)$,
$\sigma_z = \left(\begin{smallmatrix} 1 &\phantom{-}0 \\ 0 &-1\end{smallmatrix}\right)$
be the Pauli matrices.
In 
$n$ spins"~$\tfrac{1}{2}$,
a $\sigma_{kx}$ for spin $k$ is tacitly embedded as
$\unity\otimes\cdots\unity\otimes\sigma_{x}\otimes\unity\otimes\cdots\unity$
where $\sigma_{x}$ is at position $k$.
The same holds for $\sigma_{ky}$,  $\sigma_{kz}$, and in the weak coupling
terms $\sigma_{kz}\sigma_{\ell z}$ with $1\leq k<\ell\leq n$.

Now a system of $n$ 
qubits is {fully controllable} \cite{TOSH-Diss},
if e.g. the control Hamiltonians {$H_j$} comprise the Pauli matrices
$\{{\sigma_{kx}, \sigma_{ky}}\,|\, k=1,2,\dots n\}$
on every single qubit selectively and the drift Hamiltonian {$H_d$}
encompasses the Ising pair interactions
$\{{J_{k\ell}} \; {(\sigma_{kz}\sigma_{\ell z})/2}\,|\, k<\ell=2,\dots n\}$,
where the coupling topology of $J_{k\ell}\neq 0$
may take the form of {any connected graph}.
\noindent
This theorem has meanwhile been generalised to other coupling types \cite{GA02,AlbAll02}.

\medskip
\noindent
In view of the compilation task in quantum computation we get the following 
synopsis:
\begin{corollary}
The following are equivalent:
\begin{enumerate}
\item[(1)] in a quantum system of $n$ coupled spins-$\tfrac{1}{2}$, the drift $H_d$ and the controls $H_j$
        form a generating set of $\mathfrak{su}(2^n)$;
\item[(2)] the quantum system is operator controllable (in the sense of Ref.~\cite{AA03});
\item[(3)] every unitary transformation $U\in SU(2^n)$ can be realised by that system;
\item[(4)] there is a set of {universal quantum gates} for the quantum system.
\end{enumerate}
\end{corollary}
{\bf Proof:}\quad The equivalence of (1) and (2) relies on the unitary group being a compact connected
Lie group: compact connected Lie groups have no closed {\em subsemigroups} 
that are no groups themselves \cite{JS72}. 
Moreover, in compact connected Lie groups the exponential mapping is surjective, hence
(1) $\Rightarrow$ (3). Assertions
(3) and (4) just re-express the same fact in different terminology.  $\hfill\blacksquare$



\subsubsection*{Scope and Organisation of the Paper}
The purpose of this paper is to show that optimal control theory can be put to good use 
for devising multi-qubit building blocks designed for scalable quantum computing in
realistic settings. Note 
these building blocks are no longer meant to be universal {\em in the practical sense} 
that any arbitrary quantum module should be built from them (plus local controls). 
Rather they provide specialised sets of complex instructions
tailored for breaking down typical tasks in quantum computation
with substantial speed gains compared to the standard
compilation by decomposition into one-qubit and two-qubit gates.
Thus a \cisc quantum compiler translates into significant progress in fighting
relaxation. 

For demonstrating quantum \cisc compilation and scalable assembly, 
in this paper we choose systems with linear
coupling topology, i.e., qubit chains coupled by nearest-neighbour Ising interactions.
The paper is organised as follows: 
\cisc quantum compilation by optimal control will be illustrated in three different, yet
typical examples
\begin{enumerate}
\item[(1)] the indirect $1,n$-\swap gate,
\item[(2)] the quantum Fourier transform (\qft) ,
\item[(3)] the generalisation of the \cnot and Toffoli gate to multiply-controlled
		\gnot gates, \cnnot.
\end{enumerate}
For every instance of $n$-qubit systems, we analyse the effects of (i) sacrificing universality by
going to special instruction sets tailored to the problem, (ii) extending pair
interaction gates to effective multi-qubit interaction gates, and (iii) we compare the time gain
by recursive $m$-qubit \cisc-compilation ($m\leq n$) to the two limiting cases of the 
standard \risc-approach ($m=2$) on one hand and the (extrapolated) time-complexity inferred
from single-\cisc compliation (with $m=n$).

\subsection*{Preliminaries}

\subsubsection*{Time Standards}
When comparing times to implement unitary target gates by 
the \risc\/ {\em vs} the \cisc approach, 
we will assume for simplicity that local unitary operations are \/`infinitely\/' fast
compared to the duration of the Ising coupling evolution so that the total
gate time is solely determined by the coupling evolutions unless stated otherwise.
Let us emphasise, however, this stipulation only concerns the time standards. 
The optimal-control assisted \cisc-compilation methods presented here are in no way
limited to fast local controls. In particular, also the assembler step of concatenating
the \cisc-building blocks is independent of the ratio of times for
local operations {\em vs} coupling interactions.

\subsubsection*{Overview on Gate and Time Complexities}
For practical purposes, the complexity of a unitary quantum operation can be expressed in terms of two measures:
the {\em gate complexity} counts the number of universal one- and two-qubit gates
for exactly implementing the target operation in a circuit. Moreover, in view of fighting relaxation,
we will estimate the {\em time complexity} in terms of consecutive time-slots with simultaneous
$m$-qubit modules required. 

In order not to raise false expectations, upon changing from universal
2-qubit decompositions (\risc) to $m$"~qubit \cisc-implementations 
the gate complexity for exact implemention of a {\em generic} $n$"~qubit
unitary operation clearly remains {\sc np}:
it requires \/`exponentially many\/' $2$-qubit modules or
$m$-qubit modules ($m\geq 2$) alike, 
yet a cut from the order of roughly $4^n/4^2$ necessary
$2$-qubit modules down to some $4^n/4^m$ $m$-qubit modules (with up to $m=10$) 
is substantial and particularly valuable in few-qubit systems. 
More elaborate estimates will be given shortly. ---
Likewise, also in target modules with linear 2-qubit \risc complexity,
$m$-qubit \cisc complexity remains linear, yet when translated into time complexity
it may entail sizeable speed-ups -- we will show examples where they allow for
accelerations by more than a factor of $13$.
\begin{figure}[Hb!]
\begin{center}
\includegraphics[width=.45\textwidth]{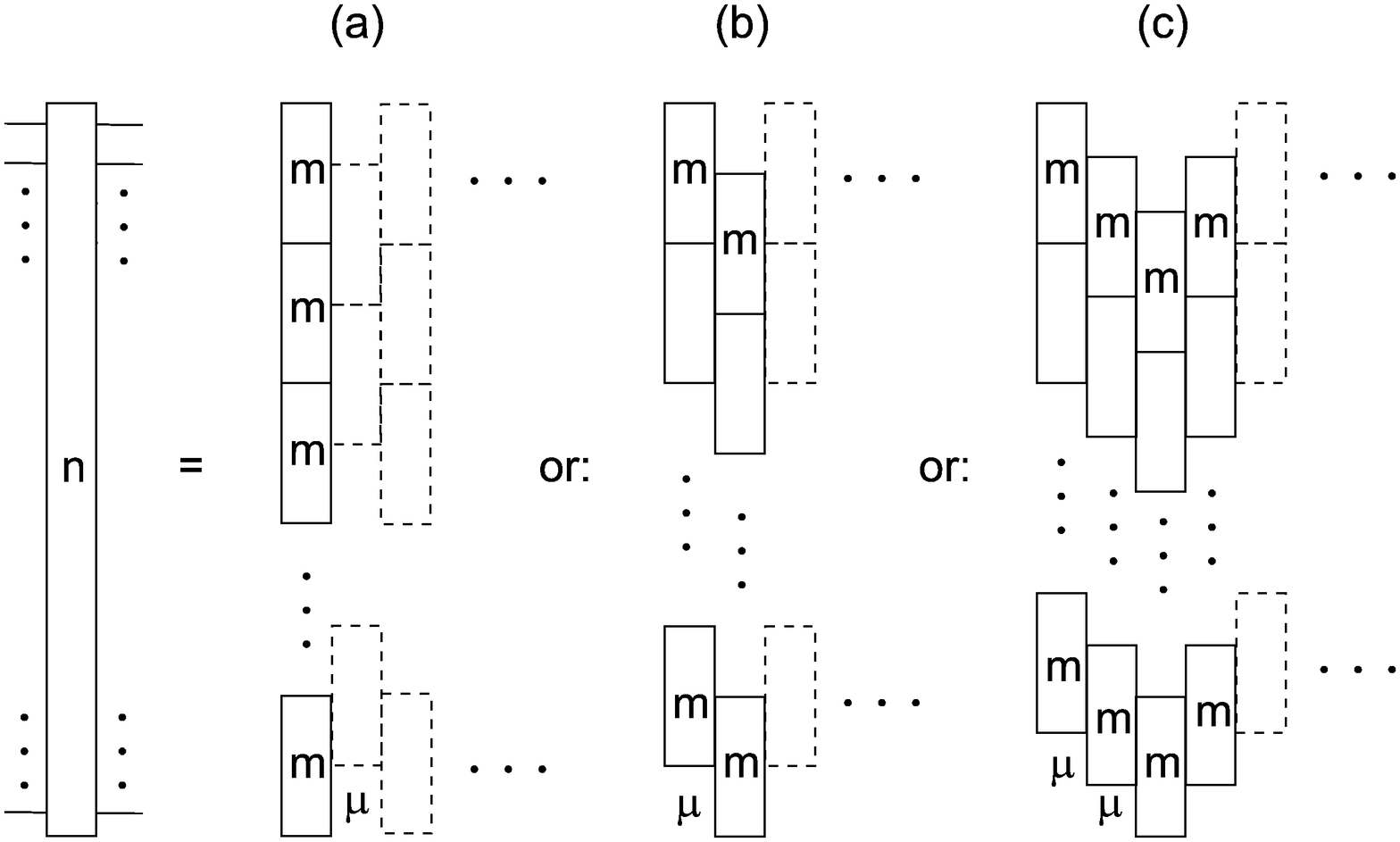}
\end{center}
\caption{\label{fig:network1}
Decomposition of an $n$"~qubit gate into a circuit of $m$"~qubit gates, where $m$ is a uniform
block size and may consist of \risc modules $m=2$ or \cisc modules with $m>2$.
(a) Margolus pattern with $\tfrac{n}{m}$ integer, (b) $n-m \floor{\tfrac{n}{m}} = \mu$ or
(c) $n-m \floor{\tfrac{n}{m}} = 2 \mu$, so $\mu>0$ integer.
}
\end{figure}

To be more precise,
a lower bound for the number of two-qubit gates necessary to exactly implement a
a generic $n$"~qubit unitary target module was given by Barenco et al.~\cite{Barenco}.
Their parameter-counting argument is based on a gem, which deserves to be picked up
for generalising it to realisations by $m$-qubit modules as illustrated in 
Fig.~\ref{fig:network1}. The key is that only in the first time slot the number
of parameters directly relates to the unitary group, while from the second slot onwards
the parameters have to be counted in terms of {\em cosets} 
of the form $SU(2^m)/(SU(2^{m-\mu})\otimes SU(2^\mu))$, if the $m$"~qubit module has
overlaps of $\mu$ qubits and $(m-\mu)$ qubits with the two adjacent modules in the
time slot before. The number of real parameters (denoted by $\#$ for short)
in the respective basic building bocks amount to
\begin{eqnarray}
\# SU(2^m) &=& 4^m - 1\label{eqn:cosets-1}\\[3mm]
\# \frac{SU(2^m)}{SU(2^{m-\mu})\otimes \unity_{2^\mu}} &=& 4^m - 1 -(4^{m-\mu} -1)\notag \\[-1mm]
		&=& 4^{m-\mu} (4^\mu -1)\quad\label{eqn:cosets-2}\\[2mm]
\# \frac{SU(2^m)}{SU(2^{m-\mu})\otimes SU(2^\mu)} &=& 4^m - 1 -(4^{m-\mu} -1) 
						- (4^\mu -1)\hspace{-9mm}\notag\\[-1mm]
		&=& (4^{m-\mu}-1) (4^\mu -1)\quad\label{eqn:cosets-3}.
\end{eqnarray}

\begin{table*}[Ht!]
\caption{\label{tab:cisc-overview}
Lower Bounds to Gate Complexities and Time Complexities for Implementing Generic $n$-Qubit Unitaries $SU(2^n)$
}
\begin{ruledtabular}
\begin{tabular}{l  c c c c c c c c  c c c}
\\[-2mm]
$n$-qubit operation with $n=$ & $2$ & $3$ & $4$ & $5$ & $6$ & $7$ & $8$ & $9$ & $10$ & $20$ & $100$\\[1mm]
\hline\\[-1mm]
no of 2-qubit gates ($g_2$) & $1$ & $6$ & $27$ & $112$ & $453$ & $1,818$ & $7,279$ & $29,124$ & $116,505$ & $1.22 \times 10^{11}$ & $1.79 \times 10^{59}$\\[1mm]
no of 2-qubit time slots ($t_2$)  & $1$ & $6$ & $14$ & $ 56$ & $151$ & $ 606$ & $1,820$ & $ 7,281$ & $ 23,301$ & $1.22 \times 10^{10}$ & $3.57 \times 10^{57}$\\[1mm]
\hline\\[-1mm]
no of 10-qubit gates ($g_{10}$) & & & & & & & & & $1$ & $1,050,627$ & $1.54 \times 10^{54}$\\[1mm]
no of 10-qubit time slots ($t_{10}$) & & & & & & & & & $1$ & $525,314$ & $1.54 \times 10^{53}$\\

\end{tabular}
\end{ruledtabular}
\end{table*}

With these stipulations one may readily determine the number of $m$"~qubit gates in a unitary network
of the type of Fig.~\ref{fig:network1}~a, where ${\frac{n}{m}}$ is integer, such as to ensure
to exhaust the number $4^n -1$ of parameters of a generic $n$"~qubit target gate
to be implemented. 
In the first time slot there are ${\frac{n}{m}}$ parallel $m$"~qubit gates (counting
by the number of parameters in the group according to Eqn.\ref{eqn:cosets-1}), in the second
time slot there are $({\frac{n}{m}}-1)$ parallel $m$"~qubit gates. They contribute
the number of parameters of the coset (Eqn.~\ref{eqn:cosets-3}), 
where one is forced to choose $\mu=\tfrac{m}{2}$ for even $m$
and $\mu=\tfrac{1}{2}(m\pm 1)$ for odd $m$ in order to be efficient. 
Following the same Margolus pattern one adds as
many $m$"~qubit gates (counting cosets) as required to superseed $4^n-1$ parameters. 
Using Gauss' brackets one thus obtains the number $g_m $ of $m$"~qubit 
gates needed to implement a generic $n$"~qubit target gate 
\begin{equation}\label{eqn:g-int}
g_m = \Big\lceil\; 
	{ \frac{4^n - 1 - {\frac{n}{m}} (4^m-1)}{4^m - 4^\mu - 4^{m-\mu} +1} 
		+ {\frac{n}{m}} }
	\;\Big\rceil
\end{equation}
and the respective number of time slots $t_m = \ceil{\tfrac{g_m}{\floor{\frac{n}{m}}}}$ by
\begin{equation}
t_m = 1+ \Big\lceil\; 
	{ \frac{4^n - 1 - {\frac{n}{m}} (4^m-1)}{{\frac{n}{m}} (4^m - 4^\mu - 4^{m-\mu} +1)} }
	\;\Big\rceil\quad.
\end{equation}
For even $n$ with $m=2$ and $\mu=1$ Eqn.~\ref{eqn:g-int} specialises to reproduce the result 
of Ref.~\cite{Barenco}, i.e. $g_2=\tfrac{1}{9}(4^n-3n-1)$.

Next, consider Fig.~\ref{fig:network1}~b and its Margolus pattern with one overhead 
of $\mu=n-m \floor{\frac{n}{m}}$ qubits to be taken into account by Eqn.~\ref{eqn:cosets-2}.
Then the same arguments give 
\begin{equation}
g'_m = \Big\lceil\; {\frac{4^n - 4^{m-\mu} - \floor{\frac{n}{m}} (4^m-1) }{4^m - 4^\mu - 4^{m-\mu} +1} 
		+ \floor{\frac{n}{m}} }\;\Big\rceil
\end{equation}

\begin{equation}
t'_m = 1+ \Big\lceil\; {\frac{4^n - 4^{m-\mu} - \floor{\frac{n}{m}} (4^m-1) }%
		{\floor{\frac{n}{m}} (4^m - 4^\mu - 4^{m-\mu} +1)} } \;\Big\rceil\;.
\end{equation}
Finally, for a pattern with two such overheads as in Fig.~\ref{fig:network1}~c,
where $n-m \floor{\frac{n}{m}}=2\mu$, one likewise finds 
\begin{equation}
g''_m = \Big\lceil\; {\frac{4^n +1 - 2\cdot 4^{m-\mu}-\floor{\frac{n}{m}}(4^m-1)}{4^m - 4^\mu - 4^{m-\mu} +1} 
		+ \floor{\frac{n}{m}} }\;\Big\rceil\qquad
\end{equation}
\begin{equation}
t''_m = 1+ \Big\lceil\; {\frac{4^n +1 - 2\cdot4^{m-\mu} - \floor{\frac{n}{m}} (4^m-1) }%
		{\floor{\frac{n}{m}} (4^m - 4^\mu - 4^{m-\mu} +1)} } \;\Big\rceil\;.\quad
\end{equation}
With efficient implementations requiring $\mu$ to be closest to $m/2$ ({\em vide supra}),
three overheads do not occur.

Since $g_m\geq g'_m\geq g''_m$, one may use $g''_m$ with the most efficient setting of
$\mu=\tfrac{1}{2}m$ for $m$ even 
or $\mu=\tfrac{1}{2}(m-1)$ for $m$ odd as a lower bound for the number of unitary
$m$"~qubit modules necessary to exactly implement an arbitrary generic $n$"~qubit target unitary.

In the limit of large $n$, one thus obtains the bounds on gate complexities
$\bar g_2\simeq 4^n/9$ and $\bar g_{10}\simeq 4^n/1,046,529$ so 
$\bar g_{10}/\bar g_2 \simeq 1/116,281 $. Likewise the limiting time complexities 
$\bar t_2\simeq 2\cdot 4^n/(n\cdot 9)$ and $\bar t_{10}\simeq 10\cdot 4^n/(n\cdot 1,046,529)$ give
a speed-up potential of $\bar t_{10}/\bar t_2 \simeq 1/23,256$ in units of the ratio of single-gate
times ${\tau_{10}}/{\tau_2}$ in the respective experimental setting. These limiting speed-up
ratios are nearly reached already for $n=10$, as the numbers given in Tab.~\ref{tab:cisc-overview}
show. In this sense, accelerations may be taken as roughly constant over the entire range of interest.

Although in generic $n$"~qubit unitaries, the \cisc speed-up may appear overwhelming, quantum algorithms
are usually by construction resorting to highly non-generic unitary bulding blocks, many of which
with linear complexities \cite{Barenco}. However, in these seemingly less rewarding yet practically
relevant cases \cisc compilation will turn out to be highly advantageous as demonstrated in three 
worked examples in the current study. --- Since generic and thus highly entangled
states have recently turned out to be computationally of modest use \cite{GrFlEi08}, recasting the above
analysis in terms of $2$"~designs and $t$"~designs \cite{Dankert06,GrAuEi07,Ambainis07} 
and following concentrations of measure will give a more realistic estimate,
which is part of a different project.

\begin{figure*}[Ht!]
{\sf \hspace{-54mm} (a) \hspace{85mm} (b)}
\begin{center}
\includegraphics[width=.45\textwidth]{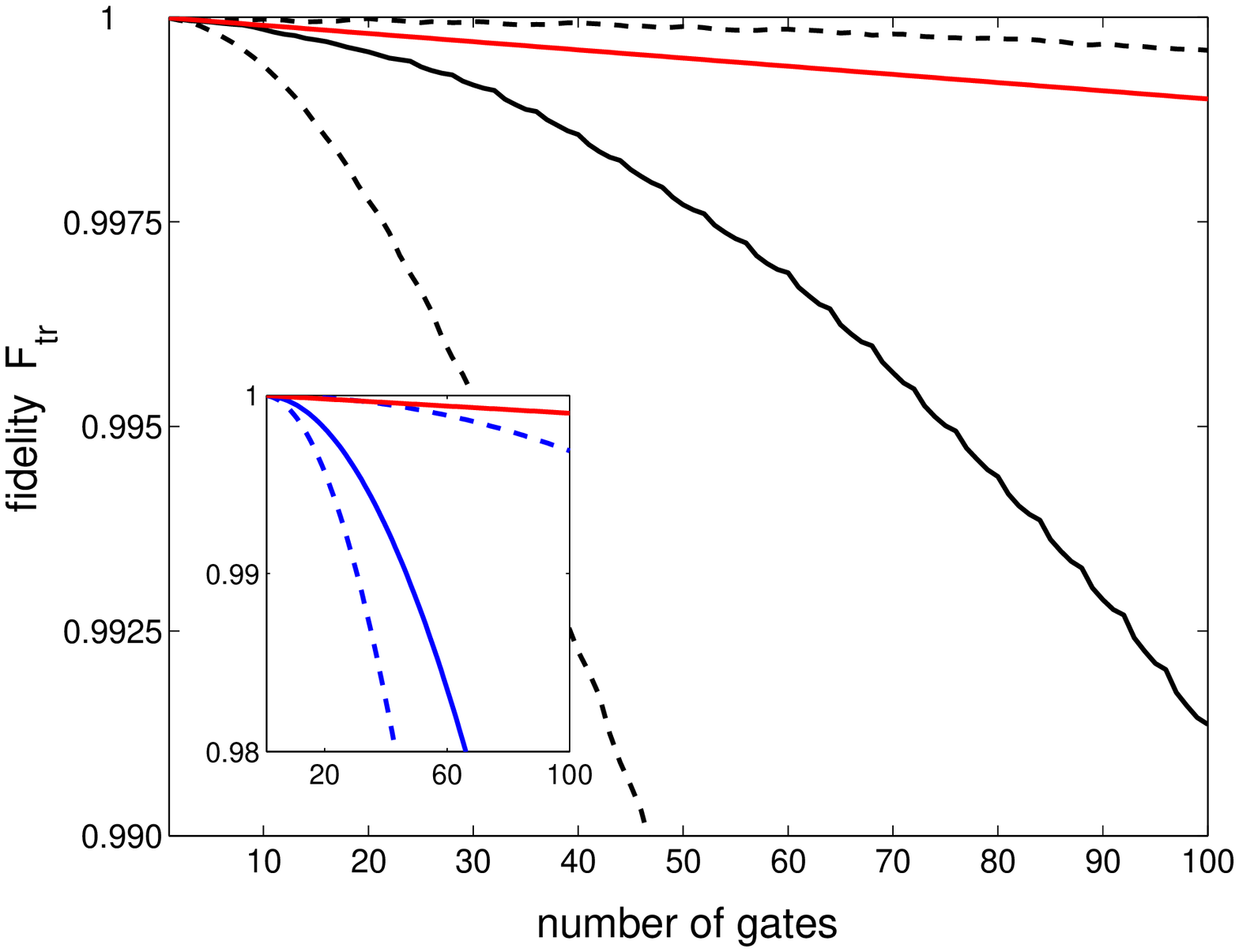}\hspace{10mm}
\includegraphics[width=.45\textwidth]{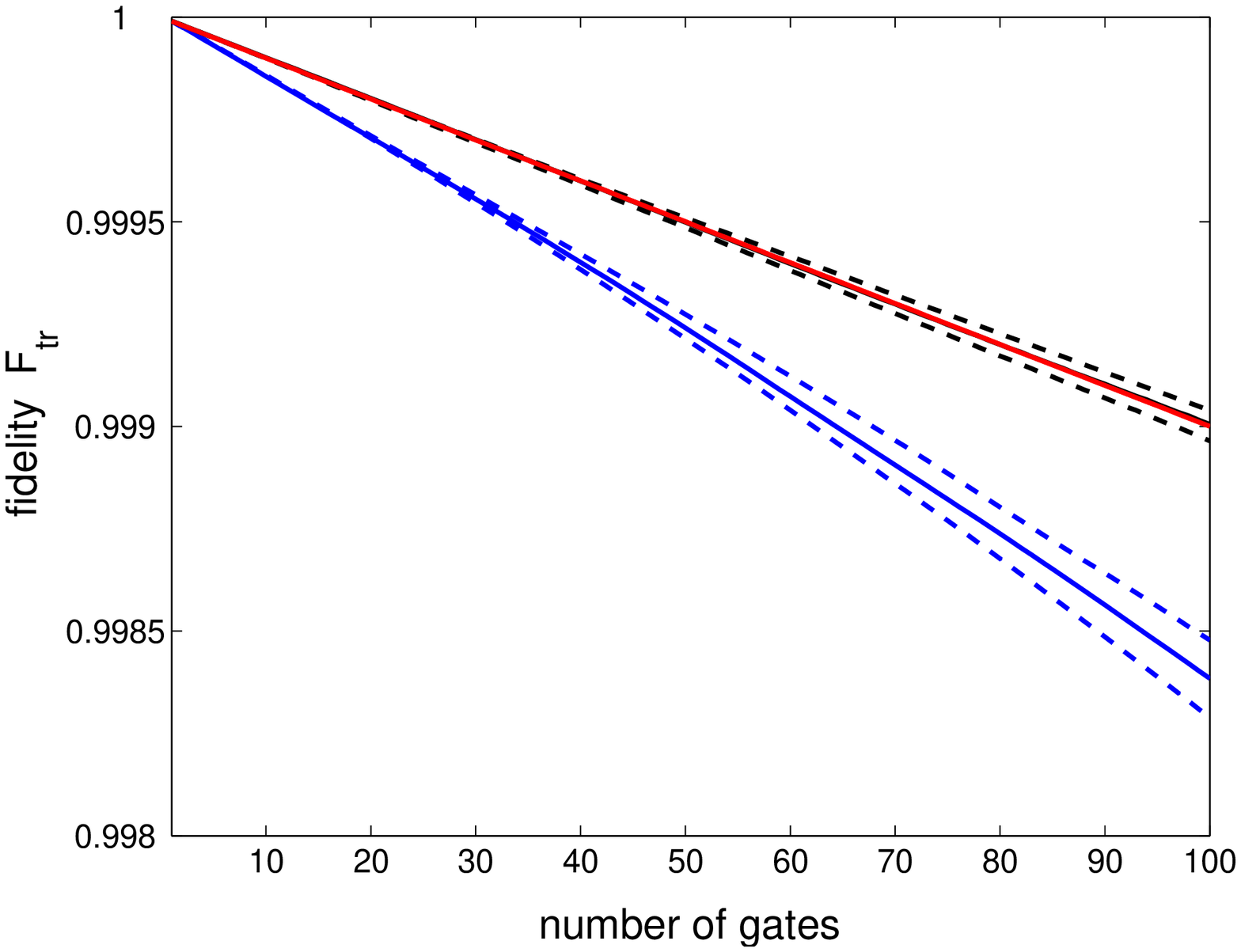}
\end{center}
\caption{\label{fig:aaaa_abcd} 
(Colour online)
Comparison of error-propagation models for random unitary gates with $m=2$ qubits (a) and $m=8$ qubits (b) 
requiring representations with different scales.
Single gate fidelity in the Monte Carlo simulations is $F_m=0.99999$. 
Repetition of the same gate $A$ (blue) is compared with repetitions
of a sequence of four independent gates $ABCD$ (black). Out of 10 Monte Carlo simulations (details see text), 
the median (solid lines) as well as the best and worst cases (dashed lines) are given. The red solid lines
denote independent error propagation $F_{\rm tr} = (F_m)^r$. Large systems ($m=8$) with several gates ($ABCD$)
resemble independent error propagation almost perfectly, as in (b) the black and the red solid lines virtually
coincide.
}
\end{figure*}
\subsubsection*{Error Propagation and Relaxative Losses}

As the main figure of merit we refer to a quality function 
\begin{equation}
q := F_{\rm tr}\; e^{-\tau/T_R}
\end{equation}
resulting from the fidelity $F_{\rm tr}$
and the relaxative decay with overall relaxation rate constant $T_R$ during a duration $\tau$ 
assuming independence of fidelity and decay. 
Moreover, for $n$ qubits one defines as the trace fidelity of an 
experimental unitary module $U_{\rm exp}$ with respect to the target gate $V_{\rm target}$ 
the quantity
\begin{equation}
\begin{split}
F_{\rm tr} &:=\; \tfrac{1}{N}\; \Re\; \tr \{V^\dagger_{\rm target} U^{\phantom{\dagger}}_{\rm exp}\}\\[2mm]
       &\phantom{:}=\; 1 - \tfrac{1}{2N}\; \fnormsq{V_{\rm target} - U_{\rm exp}}\quad,
\end{split}
\end{equation}
where both $U,V\in U(N)$ with $N:=2^n$.
It follows via the simple relation to the Euclidean distance
\begin{equation*}
\begin{split}
\fnormsq{V - U} &= \fnormsq{U} + \fnormsq{V} - 2 \Re \tr \{V^\dagger U\}\\[2mm]
		&= 2 N - 2 N \tfrac{1}{N} \Re \tr \{V^\dagger U\}\\[2mm]
		&= 2 N (1 - F_{\rm tr})\quad,
\end{split}
\end{equation*}
the latter two identities invoking unitarity of $U,V$. The reason for chosing the trace fidelity
is its convenient Fr{\'e}chet differentiability in view of gradient-flow techniques, 
see also Ref.~\cite{SGDH08}.

Consider an $m$"~qubit-interaction module (\cisc) with quality $q_m = F_m\;e^{-\tau_m/T_m}$ that decomposes into
$r$ universal two-qubit gates (\risc), out of which $r\,'\leq r$ gates have to be performed {\em sequentially}.
Moreover, each 2"~qubit gate shall be carried out with the uniform quality $q_2 = F_2\;e^{-\tau_2/T_2}$. 
Henceforth we assume for
simplicity equal relaxation rate constants, so $T_2 = T_m$ are identified with $T_R$. 
Then, as a first useful rule of the thumb and assuming independent error propagation, 
it is advantageous to compile the $m$"~qubit module directly
if $F_m > (F_2)^r$. Or more precisely taking relaxation into account, 
if the module can be realised with a fidelity
\begin{equation}
F_m > (F_2)^r \; e^{-(r\,'\cdot\tau_2 - \tau_m)/T_R}\quad.
\end{equation}

A more refined picture emerges from Monte-Carlo simulations of error propagation.
To this end, compare the above independent error estimates with
two scenarios for a sequence of $r$ gates in total: 
(i) the $r$"~fold repetition of single unitary gates $A$
with individual errors meant to give $A^r$ with $r=1,2,3,\dots$ and 
(ii) the repetition of a sequence of four different gates $A,B,C,D$
again each with individual errors
to give $(D\circ C \circ B \circ A)^{r/4}$ where $r=4,8,12,\dots$. 
In the sequel, we refer to case (i) as $AAAA$ and to case (ii) as $ABCD$.

\begin{figure*}[Ht!]
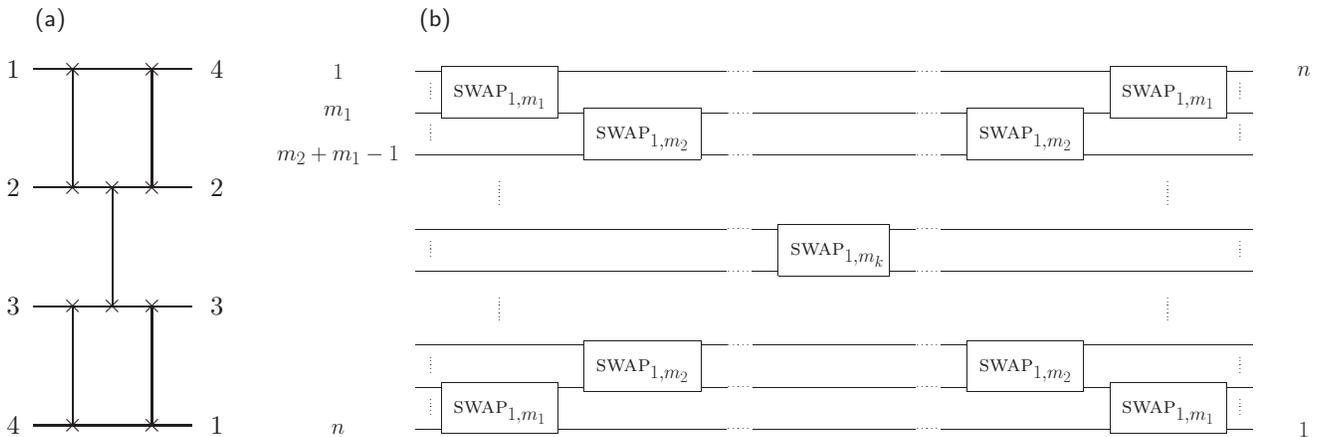

\mbox{\sf\hspace{-14mm}(a)\hspace{47mm}(b)\hspace{55mm}\hspace{42mm}}\\[3mm]
\mbox{
\includegraphics[height=.212\textheight]{recSWAP1net.epsi}
\hspace{.5cm}
\includegraphics[height=.212\textheight]{recSWAP2net.epsi}
}
\caption{\label{fig:SWAPnet} (a) Simple starting point: building a \swapj 4 gate from 
five \swapsj 2.
(b) Generalisation: assembling a \swapj n by four \swapsj {m_j} for each type $j=1,2,\dots,k-1$
and one single \swapj {m_k} so that $m_k + 2 \sum_{j=1}^{k-1} (m_j -1) = n$. 
}
\end{figure*}
For gates and errors to be generic, we use random unitaries (distributed according 
to the Haar measure following a recent modification \cite{Mez07} of the {\sc qr}"~algorithm).
To a given random unitary $m$"~qubit gate 
$A_0 \in U(2^m)$ (defining its Hamiltonian $H_{A0}$ via $A_0=e^{-i H_{A0}}$) 
we simulate a generic error as follows: from another
independent unitary $E_j$ take the matrix logarithm $H_{Aj}$ such that $e^{-i H_{Aj}}= E_j$.
Then to a given trace fidelity $F$, a corresponding unitary with a Monte-Carlo random error 
(the error being introduced on the level of the Hamiltonian generators) can readily be
obtained by solving
\begin{equation}
\begin{split}
F &= 1 - \frac{1}{2N} ||A_0 - A_j||_2^2 \\[2mm]
	&= 1 - \frac{1}{2N} \big|\big|A_0 - e^{-i(H_{A0} + \delta\cdot H_{Aj})}\big|\big|_2^2
\end{split}
\end{equation}
for $\delta>0$.
Along these lines one obtains the Monte-Carlo fidelities for repeating the $A$"~gate
by
\begin{equation}
F_{AAAA}(r) = 1 - \frac{1}{2N} \big|\big|(A_0)^r - \prod\limits_{j=1}^r A_j \big|\big|_2^2
\end{equation}
and
\begin{equation}
F_{ABCD}(r) = 1 - \frac{1}{2N} \big|\big|(D_0 C_0 B_0 A_0)^{\tfrac{r}{4}} 
		- \prod\limits_{j=1}^{\tfrac{r}{4}} (D_j C_j B_j A_j) \big|\big|_2^2\;,
\end{equation}
where the product runs from right to left.
These Monte-Carlo simulations are compared to the simple model of independent errors
according to 
\begin{equation}
F_{\rm ind} = (F_m)^r\quad .
\end{equation}
As shown in Fig.~\ref{fig:aaaa_abcd}~a, for two-qubit gates the error propagates with a vast
variance, which makes it virtually unpredictable. Thus assuming independence is
always too optimistic for AAAA, while for ABCD it is still mostly optimistic,
although there are cases in which the errors may compensate to give less effective
loss than expected under independence.

However, when moving to effective multi-qubit gates, i.e., \cisc modules,
the generic situation becomes more predictable. For example, in $8$"~qubit random unitary gates,
Fig.~\ref{fig:aaaa_abcd}~b shows that AAAA is significantly deviating from independent error
propagation, whereas ABCD resembles independent error propagation almost perfectly.
The situation is qualitatively exactly the same even if the single gate error is larger
as tested by analogous Monte-Carlo simulations setting 
$F_{\rm tr} = 0.99$ or $F_{\rm tr} = 0.96$ (not shown).

In the sequel, we will---for the sake of 
simplicity---often assume independent error propagation at the expense of systematically
underestimating the pros of \cisc compilation compared to the standard \risc compilation into universal
local and two-qubit gates.

\subsubsection*{Computational Methods and Devices}
Following the lines of our previous work on time complexity \cite{PRA05},
we used the \grape algorithm \cite{GRAPE} for direct \cisc compilation.
It tracks the fixed final times down to the shortest durations of controls
still allowing for synthesising the unitary target gates with full fidelity. 
This gives currently the best known upper bounds to the minimal times required
to realise a target module on a concrete hardware setting.
We extended our parallelised {\sc c++} code of the \grape package described in \cite{EP06}
by adding more flexibility allowing to efficiently exploit available parallel nodes independent
of internal parameters \cite{HLRB07}.
Moreover, faster algorithms for matrix exponentials on high-dimensional systems based on 
approximations by Tchebychev series have been developed \cite{Wald07} specifically
in view of application to large quantum systems \cite{HLRB07}.
Thus computations could be performed on the \hlrbii supercomputer cluster
at {\em Leibniz Rechenzentrum} of the Bavarian Academy of Sciences Munich.
It provides an {\sc sgi} Altix 4700 platform
equipped with 9728 Intel Itanium2 Montecito Dual Core processors with a clock rate
of $1.6$ GHz, which give a total {\sc linpack} performance of $63.3$ TFlops/s.
The present explorative study exploited the time allowance of approx. $500.000$
\cpu hours.

\section{The $1,n$ SWAP Operation}\label{sec:iswap}
\begin{figure*}
\mbox{\sf\hspace{10mm}(a)\hspace{78mm}(b)\hspace{65mm}}\\[3mm]
\includegraphics[scale=0.4]{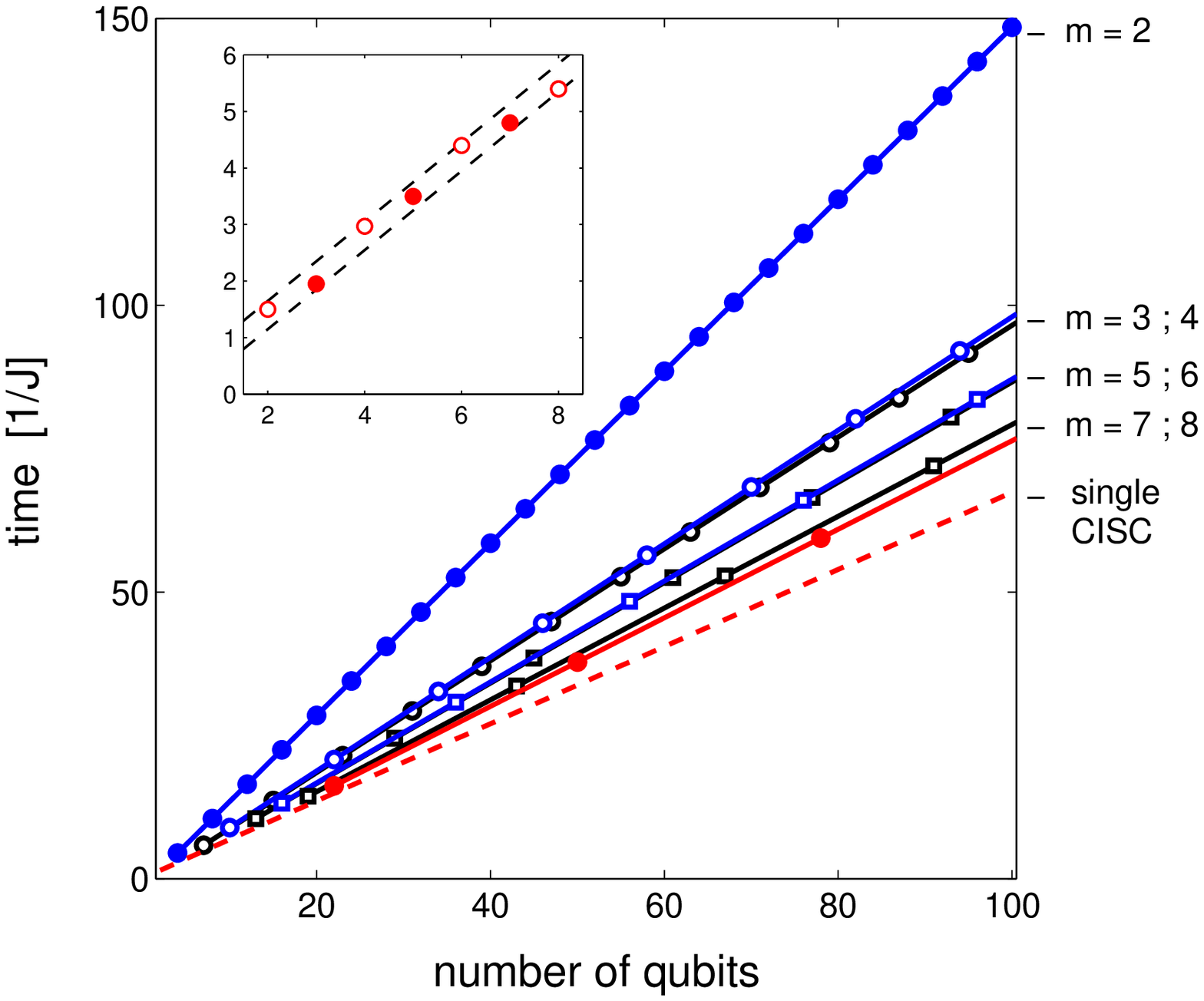}\hspace{10mm} 
\includegraphics[scale=0.4]{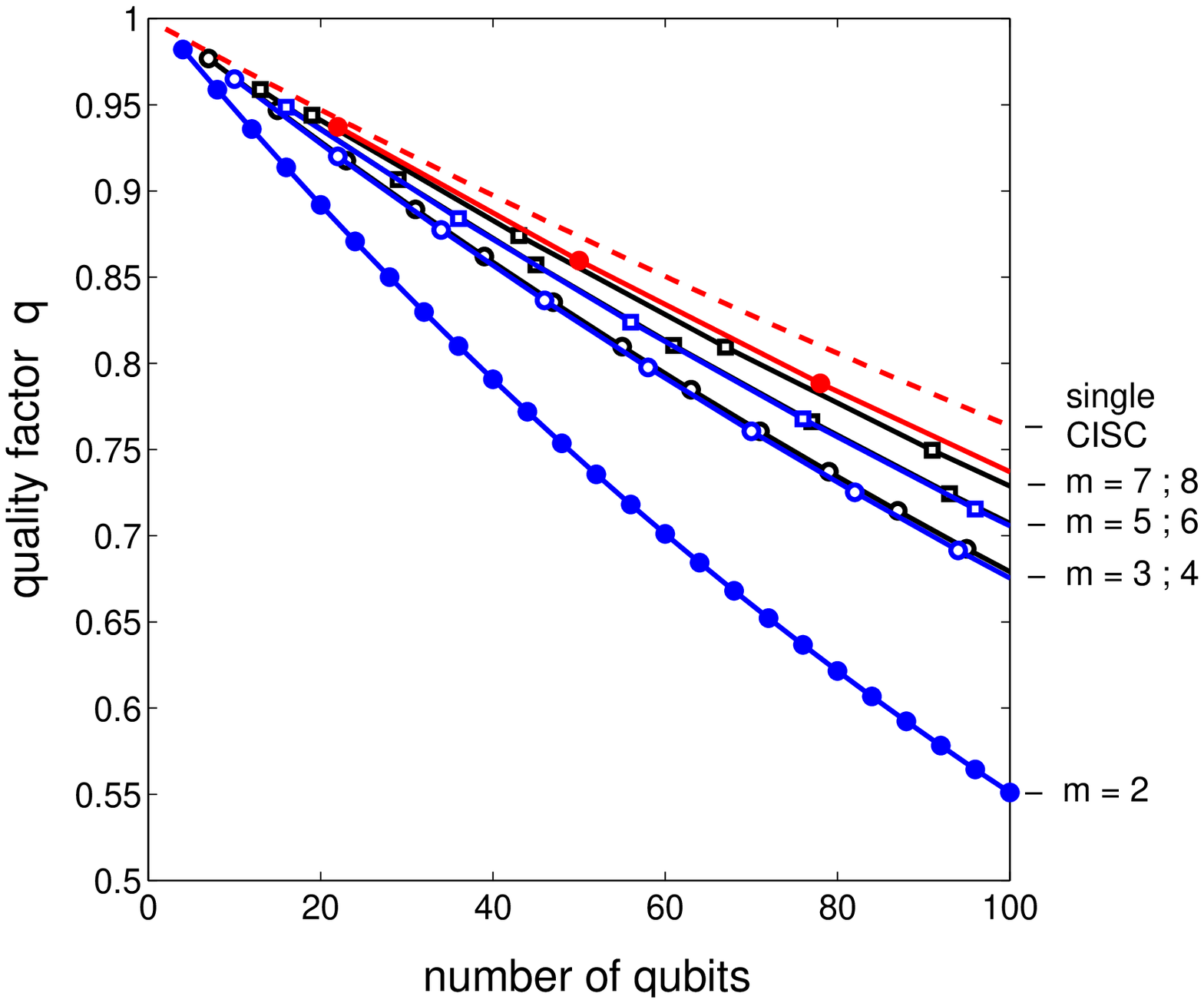}\hspace{10mm} 
\caption{\label{fig:SWAPrec} (Colour online) 
(a): Times required for indirect \swapsj n on linear chains of $n$
Ising-coupled qubits by assembling \swapj m building blocks
reaching from $m=2$ (\risc) up to $m=8$ (\cisc). Using linear regression,
the dashed line is an extrapolation
of the direct single-\cisc compilations shown in the inset to large number of qubits,
where direct \cisc compilation is virtually impossible on classical computers.
Time units
are expressed as $1/J_{ZZ}$ assuming the duration of local operations
can be neglected compared to coupling evolutions (details in the text).
(b): Translation of the effective gate times into overall quality
figures $q=(q_m)^{r_m}$ for an effective gate assembled from $r_m$ components
of single qualities $q_m:= F_m\; e^{-\tau_m/T_R}$ (with the respective component fidelities
homogeneously falling into a narrow interval $F_m \in [0.99994 , 0.99999]$ for $m=3,\dots,8$). 
Data are shown for a uniform relaxation rate constant of $1/T_R = 0.004 J_{ZZ}$.
}
\end{figure*}

The easiest and most basic examples to illustrate the
pertinent effects of optimal-control based \cisc-quantum compilation are the
respective indirect \swapj n  gates
in spin chains of $n$ qubits coupled by nearest-neighbour Ising interactions
with $J_{ZZ}$ denoting the coupling constant.

For the \swapj 2 unit there is a standard textbook decomposition
into three \cnots. Thus for Ising-coupled systems and in the limit of fast local controls,
the total time required for an \swapj 2 is $3/(2J_{ZZ})$, 
and there is no faster implementation \cite{Khaneja01b,Khaneja02}.
Note, however, that in systems coupled by the isotropic Heisenberg interaction $XXX$,
the \swapj 2 may be directly implemented just by letting the system evolve for a time
of only $1/(2J_{XXX})$. Sacrificing universality, it may thus be advantageous to 
regard the \swapj 2 as basic unit for the \swapj n task rather than the universal
\cnot. 
Clearly, any even-order \swapj {2n} can be built from \swapsj 2
along the lines of the most obvious scheme of Fig.~\ref{fig:SWAPnet}~a.
(The odd-order \swapsj {2n-1} follow, e.g., from \swapj {2n} by omitting qubit $2n$ and all the gates 
connected to it.)

Moreover, the generalisation to decomposing a \swapj n into a sequence with
$k$ different \swapj {m_j} building blocks (where $j=1,2,\dots,k$) as shown in Fig.~\ref{fig:SWAPnet}~b
is straightforward by ensuring $m_k + 2 \sum_{j=1}^{k-1} (m_j -1) = n$.
Due to its symmetry, the total duration then amounts to
\begin{equation}
\tau(\swapj n) = \tau(\swapj {m_k}) + 2 \sum\limits_{j=1}^{k-1} \tau(\swapj{m_j}) 
\end{equation}
and the overall quality as a function of the fidelities of the constitutent gates
reads
\begin{equation}
\begin{split}
q_{\swapj n} = F(\swapj {m_k}) &\quad \prod\limits_{j=1}^{k-1} F(\swapj {m_j})^4 \\
			&\times e^{-\tau(\swapj n)/T_R} \quad.
\end{split}
\end{equation}

Now, the \swapj {m_j} building blocks themselves
can be precompiled into time-optimised single complex instruction sets 
by exploiting the \grape-algorithm of optimal control up to the current
limits of $m_j$ imposed by \cpu-time allowance. 

Proceeding in the next step to large $n$, Fig.~\ref{fig:SWAPrec} underscores how 
the time required for \swapj n gates decreases significantly
by assembling precompiled \swapj {m_j} building blocks as \cisc units recursively  
up to a multi-qubit interaction size of $m_j=8$, where the speed-up is by a factor of nearly $2$. 
Clearly, such a set of \swapj {m_j} building blocks with $m_j\in\{2,3,4,5,6,7,8\}$ 
allows for efficiently synthesising any \swapj n.
Assuming for the moment that a linear time complexity of the \swapj n can be taken
for granted, one may extrapolate the results of direct \cisc compilation from the range of 
the inset of Fig.~\ref{fig:SWAPrec}~a to a large number of qubits. 
One thus obtains an estimated upper limit to the time complexity of the \swapj n. This is
indicated by the dashed line, the slope of which will be defined as $\Delta_\infty$. 
Likewise, the irrespective slopes of the $m$-qubit decomposition are denoted by $\Delta_m$.

With these stipulations, we introduce as a measure for the potential of \cisc compilation
(versus \risc compilation) the ratio of the slopes
\begin{equation}
\pi_\cisc := \frac{\Delta_2}{\Delta_\infty}
\end{equation}
and as a measure for the extent to which this potential has been exhausted by
$m$"~qubit \cisc compilation the ratio
\begin{equation}
\eta_m := \frac{\Delta_\infty}{\Delta_m}
\end{equation}
thus providing as convenient measure of improvement
\begin{equation}
\xi_m := \frac{\Delta_2}{\Delta_m} = \eta_m \cdot \pi_\cisc \quad.
\end{equation}

The data of Fig.~\ref{fig:SWAPrec} thus give a potential of $\pi_\cisc = 2.16$;
by $m=8$"~qubit interactions it is already pretty well exhausted, as inferred from $\eta_8 = 0.87$.
The current \cisc over \risc improvement then amounts to $\xi_8 = 1.88$.

On the other hand, deducing from Fig.~\ref{fig:SWAPrec} right away that the time complexity
of \swapsj n ought to be linear would be premature: although the slopes seem
to converge to a non-zero limit, numerical optimal control may become
systematically inefficient for larger interaction sizes $m$. Therefore,
although improbable, e.g.,
convergence of the slopes to a value of zero cannot be safely excluded
on the current basis of findings. This also means 
a logarithmic time complexity can ultimately not be excluded either.


\begin{figure*}[Ht!]
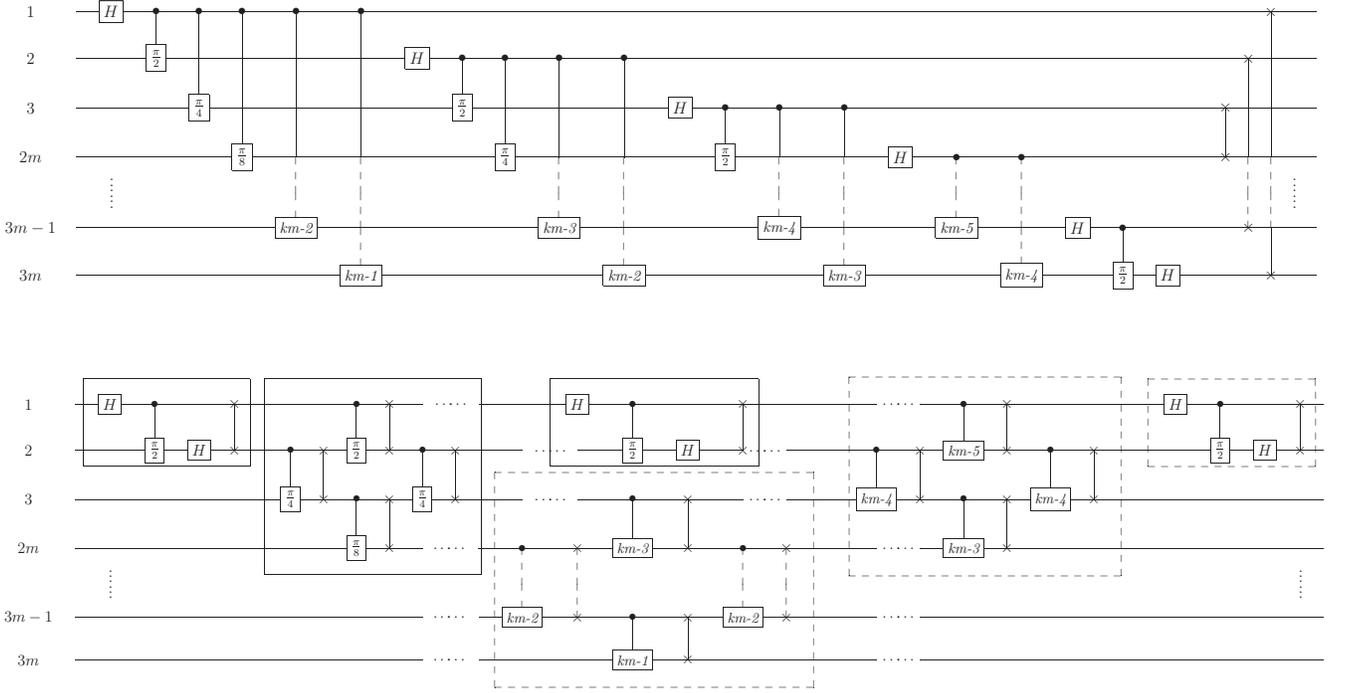

\includegraphics[scale=.523]{qft_induction1.epsi}\\[11mm]
\includegraphics[scale=.5]{qft_induction2e.epsi}
\caption{ \label{fig:qft-induction} By rearranging the \swaps and controlled phase gates,
the standard decomposition of a $3m$"~qubit quantum Fourier transform, \qft (top trace)
reduces to a realisation adapted to a coupling topology of linear nearest-neighbour interactions  (lower trace) with
a $2m$"~qubit \qft, $m$"~qubit \cpswaps (solid boxes), and an $m$"~qubit \qft (dashed box). 
The notation $(km-\nu)$ is a shorthand for a rotation angle of $\phi=\frac{\pi}{2^{km -\nu}}$.
}
\end{figure*}

\begin{figure}[Ht!]
\begin{center}
\hspace{-2mm}
\includegraphics[width=.475\textwidth]{BlockQFT_m_2b.epsi}
\caption{\label{fig:qft2-block} 
For $k\geq 2$, a $(km)$"~qubit \qft can be assembled from $k$ times an $m$"~qubit \qft and $k\choose 2$ instances of
$2m$"~qubit modules \cpswapJM j{2m}, where the index $j$ of different phase-rotation angles 
takes the values $j=1,2,\dots, k -1$.
The dashed boxes correspond to Fig.~\ref{fig:qft-induction} and show the induction $k\mapsto k+1$.
}
\end{center}
\end{figure}

Summarising the results for the indirect \swaps
in terms of the three criteria described in the introduction, we have
the following:
(i) in Ising coupled qubit chains, there is no speed-up by changing the 
basic unit from the universal \cnot into a \swapj 2, whereas in
isotropically coupled systems the speed-up amounts to a factor of three;
(ii) extending the building blocks of \swapj m from $m=2$ (\risc)
to $m=8$ (\cisc) gives a speed-up by a factor of nearly two  
under Ising-type couplings;
(iii) the numerical data are consistent with a time complexity converging to a linear limit
for the \swapj n task in Ising chains, however, there is no proof for it yet.

\section{The Quantum Fourier Transform (QFT)}

Since many quantum algorithms take advantage of efficiently
solving some hidden subgroup problem, the quantum Fourier transform
plays a central role \cite{Jozsa88, Mosca88,EHK04}.

In order to realise a \qft on large qubit systems, our approach 
is the following: given an $m$-qubit \qft, we show that
for obtaining a $(k \cdot m)$-qubit \qft by recursively
using multi-qubit building blocks, a second type of module is required,
to wit a combination of controlled phase gates and \swaps, which 
henceforth we dub $m'$"~qubit \cpswap for short.
\begin{figure*}[Ht!]
\mbox{\sf\hspace{22mm}(a)\hspace{83mm}(b)\hspace{65mm}}\\[3mm]
\begin{center}
\includegraphics[width=.40\textwidth]{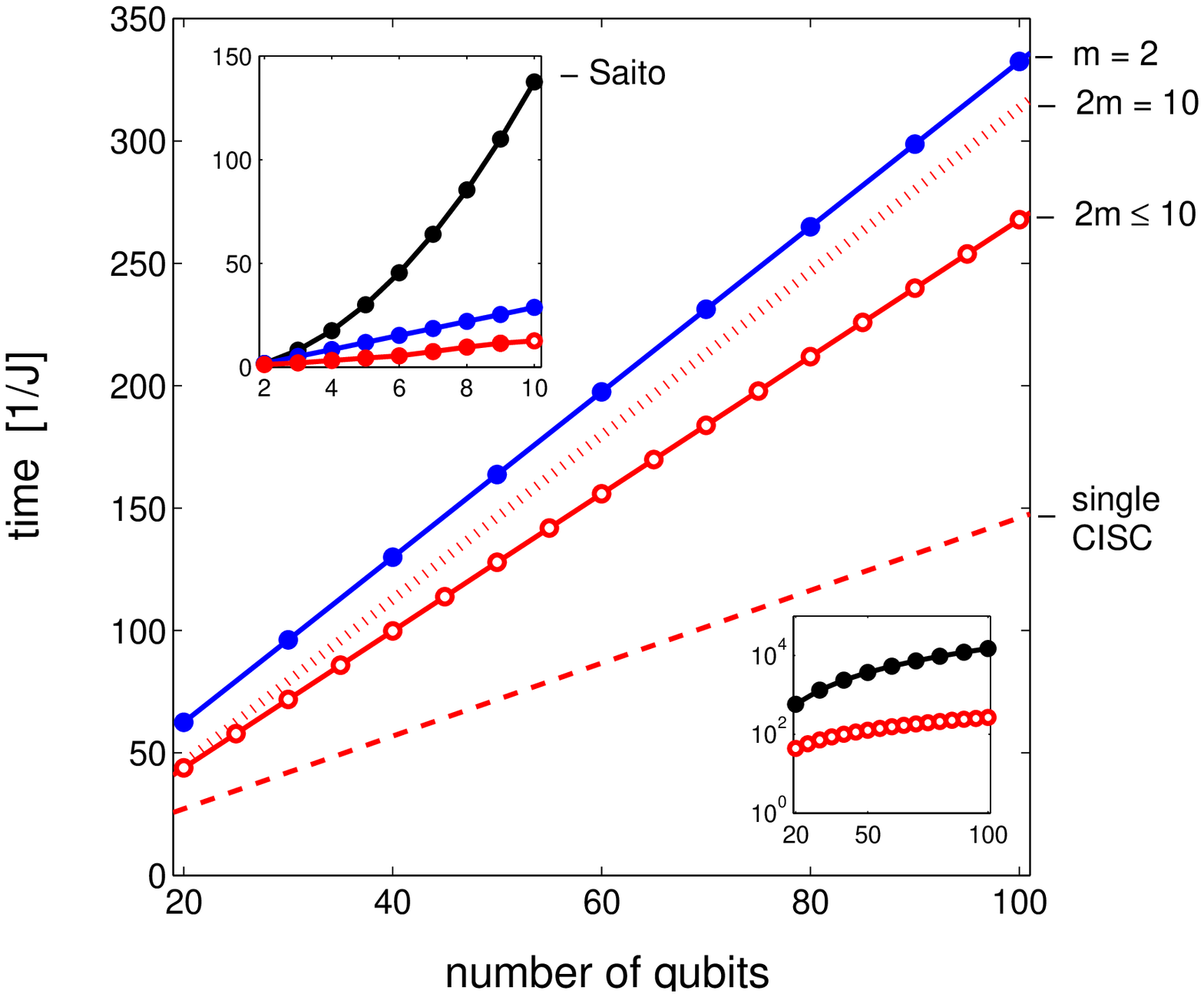}\hspace{10mm}
\includegraphics[width=.39\textwidth]{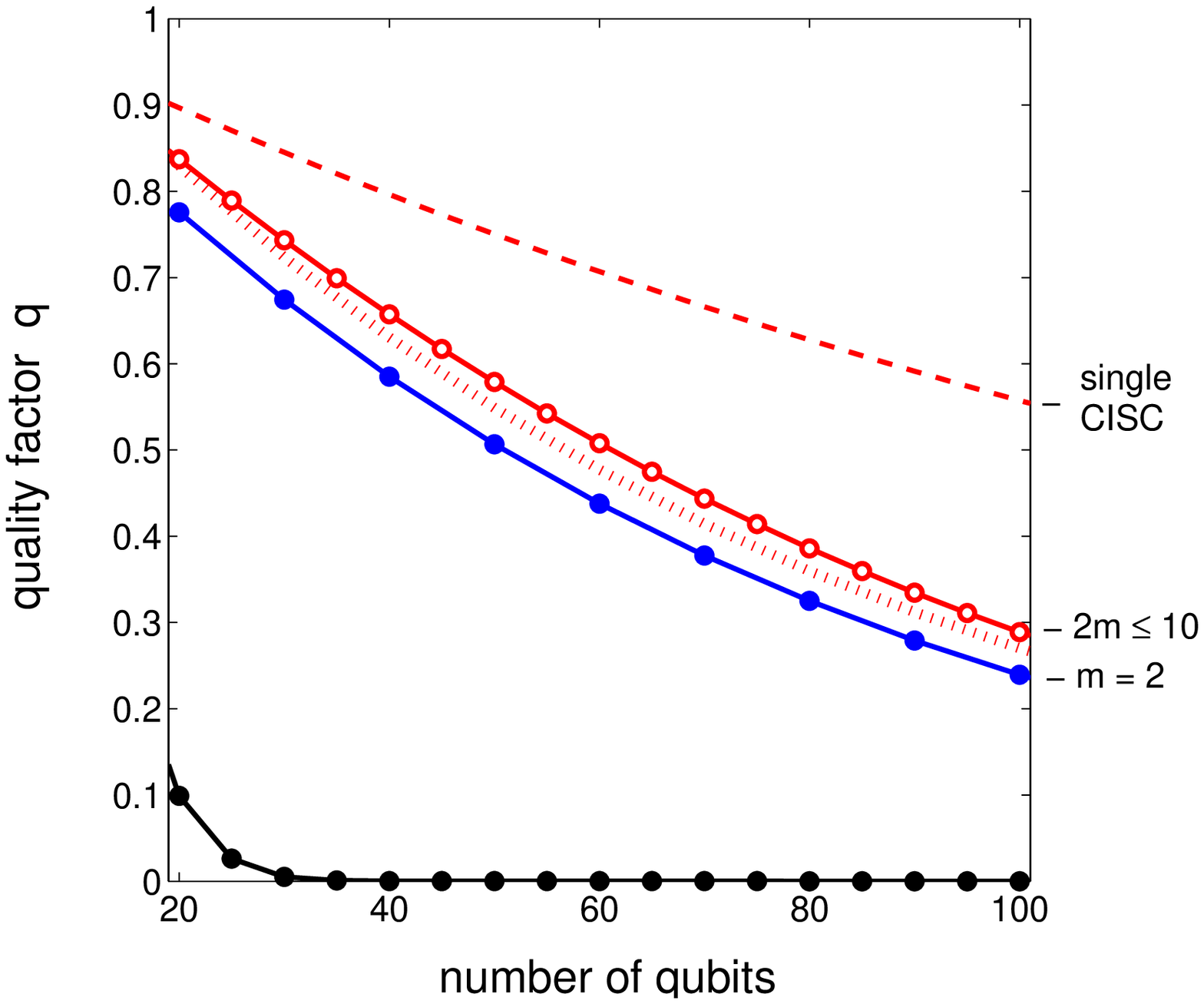}
\end{center}
\caption{\label{fig:rec_qft} (Colour online) 
Comparison of \cisc-compiled \qft (red) with standard 
\risc compilations ($m=2$) following the
scheme by Saito \cite{Saito} (black) or Blais \cite{Blais} (blue): (a) times for implementation
translate into quality factors (b) for a relaxation rate constant of $1/T_R = 0.004 J_{ZZ}$.
Again, the dashed red line extrapolates 
from the direct single-\cisc compilations shown in the 
upper inset of (a), the lower inset giving in logarithmic scales the times needed for the standard textbook
\risc compilation ($m=2$) on a linear Ising chain.
Dotted red lines represent the less favourable results from \qft variant II (Appendix A).
}
\end{figure*}

Here we present two alternatives: variant~I with $m'=2m$ and, as a special case, variant~II
for even $m'=m$.

Choosing $m=2$ and $k=3$ for a start, the recursive construction is illustrated in 
Fig.~\ref{fig:qft-induction}.
The top trace shows the standard textbook realisation of a $6$-qubit \qft. By shifting the final 
\swap operations, it can be rearranged into the sequence of gates depicted in the lower trace. 
Note that the gates appearing in solid boxes constitute a $2m$"~qubit \qft 
(which itself is made of two $m$"~qubit \qfts and a central $m$"~qubit \cpswap), 
while the ones in dashed boxes have to be added for a $3m$-qubit \qft. 
For $m=2$ we have thus shown how a $3m$"~qubit \qft reduces to a $2m$"~qubit \qft, two
$2m$"~qubit \cpswaps, and an $m$"~qubit \qft. So with $2m$ providing a foundation,
at the same time
we have also illustrated the induction from a $k\cdot m$"~\qft to a $(k+1)\cdot m$"~\qft. 
Moreover, the same construction principle holds for any block size $m=2,3,\dots$,
which can readily be proven by a straightforward, but lengthy induction from $m$ to $m+1$.

One thus arrives at the desired block decomposition of a general $(k\cdot m)$"~qubit \qft 
as shown in Fig.~\ref{fig:qft2-block} (which is variant I; the less effective 
variant II can be found in Appendix A):
it requires $k$ times the same $m$"~qubit \qft
interdispersed with $k\choose 2$ times an $2m$"~qubit \cpswap, out of which $k-1$ 
show different phase-roation angles.
For all $m$ and \mbox{$j=1,2,\dots,(k-1)$}, one finds the following observations:
\begin{enumerate}
\item a \cpswapJM j{2m} takes as least as long as a \qftm ;
\item a \qftm takes as least as long as a \cpswapjm ;
\item a \cpswapjm takes least as long as a \cpswapJM {j+1}m .
\end{enumerate}
Thus the duration of a $(k\cdot m)$"~qubit \qft built from $m$"~qubit and $2m$"~qubit
modules amounts to
\begin{equation}\label{eqn:tau-qft}
\begin{split}
        \tau(\qft_{k\cdot m}) = 2\cdot &\tau(\qft_m) + (k-1)\cdot \tau(\cpswapJM 1{2m})\\
                 &+ (k-2)\cdot \tau(\cpswapJM 2{2m})\quad.
\end{split}
\end{equation}
Next, consider the overall quality of
a $(k\cdot m)$"~qubit \qft in terms of its two types of building blocks,
namely the basic $m$"~qubit \qft as well as the constituent $2m$-qubit \cpswaps 
with their respective different rotation
angles. It reads 
\begin{equation}
\begin{split}
q_{\qft_{k\cdot m}} =\; (F_{\qft_{m}})^{k}\; 
\big(\prod\limits_{j=1}^{k-1}&\;(F_{\cpswapJM j{2m}})^{k-j}\;\big) \\
	&\qquad\times\; e^{-(\tau_{\qft_{k\cdot m}}/T_R)} \,.
\end{split}
\end{equation}
In the following, 
we will neglect rotations as soon as their angle falls short of a threshold of $\pi/2^{10}$. This
approximation is safe since it is based on a calculation of a
$20$"~qubit \qft, where the truncation does not introduce any relative error beyond $10^{-5}$.
According to the block decomposition of Fig.~\ref{fig:qft2-block}, thus three instances of \cpswaps
are left, since all \cpswapJM j{10} elements with $j\geq 3$ boil down to mere 
\swap gates due to truncation of small rotation angles. 
The representation of these \cpswap modules is shown in Appendix B as Fig.~\ref{fig:cp-swaps}.

With these stipulations, we address the task of assembling an $(k\cdot 10)$"~qubit \qft,
exploiting the limits of current allowances on the \hlrbii cluster. This translates into using
$10$"~qubit \cpswap building blocks ($2m=10$) and the $5$"~qubit \qft ($m=5$) in the sense
of a $(2k\cdot 5)$"~qubit \qft. 
Its duration $\tau(\qft_{2k\cdot 5})$ is readily obtained as in Eqn.~\ref{eqn:tau-qft} thus giving 
an overall quality of
\begin{equation}
\begin{split}
        q_{\qft_{2k\cdot 5}} =\; &(F_{\qft_{5}})^{2k} \; (F_{\cpswapJM 1{10}})^{2k-1}\;
                (F_{\cpswapJM 2{10}})^{2k-2}\\    
		&\times\;\,(F_{\cpswapJM 3{10}})^{{2k\choose 2}-4k+3}\quad  e^{-\tau_{\qft_{2k\cdot 5}}/T_R }\,.
\end{split}
\end{equation}

Based on this relation, the numerical results of Fig.~\ref{fig:rec_qft} show
that a \cisc-compiled \qft is moderately superior to the standard \risc versions \cite{Saito,Blais}.
Although the potential of \cisc compilation amounts to $\pi_\cisc= 2.27$, 
recursively assembling $5$"~qubit \qfts and $10$"~qubit \cpswaps only exploits
about half of it as apparent in the value of $\eta^{\phantom{|}}_{5,10} = 0.53$.

As has been pointed out by Zeier \cite{Robert},
the decomposition of a many-qubit \qft 
into smaller \qfts and concatenations of a permutation matrix and a diagonal matrix 
roots back in a principle already used in the Cooley-Tukey algorithm \cite{CT65} for the discrete
Fourier transform (\dft):
Let $N=m\cdot q$. Then one obtains \cite{CB93,Egner97}
\begin{equation}\label{eqn:dft_dec}
\begin{split}
\dft_N &= L \circ (\dft_m \otimes \unity_q)\circ  D\circ  (\unity_m \otimes \dft_q) \\
	&= (\unity_q \otimes \dft_m)\circ (L \circ  D) \circ  (\unity_m \otimes \dft_q)\;,
\end{split}
\end{equation}
where $L \in \Mat_N$ is a permutation matrix. Moreover, setting
$\omega := e^{2\pi i/N}$, the diagonal matrix takes the form
\begin{equation}
D = \diag(\omega^{t_k} | t_k = (k \mod m) \lfloor\tfrac{k}{m}\rfloor\; \text{for}\; k = 0,1,2,\dots N-1)
\end{equation}

Therefore, the \qft decompositions made use of here exactly follow the classical
scheme in the second line of Eqn.~\ref{eqn:dft_dec}, the expression $(L \circ  D)$
corresponding to the \cpswap.


\section{The Multiple-Controlled NOT Gate (C$^n$NOT)}

Multiply-controlled \cnot gates generalise Toffoli's gate.
Here, we move from \ctnot to \cmtnot in an $n$"~qubit system with one ancilla and one target
qubit. The reason for the ancilla qubit being that it turns the problem to linear 
complexity \cite{Barenco}. Moreover, in view of realistic large systems, we assume again a 
topology of a linear chain coupled by nearest-neighbour Ising interactions. 
Since \cmnot-gates frequently occur in error-correction schemes,
they are highly relevant in practice.

Here we address the task of decomposing a \cmtnot into lower \cnots and indirect 
\swap gates (see Sec.~\ref{sec:iswap}).

\begin{figure}[Ht!]
\begin{center}
\includegraphics[width=.4\textwidth]{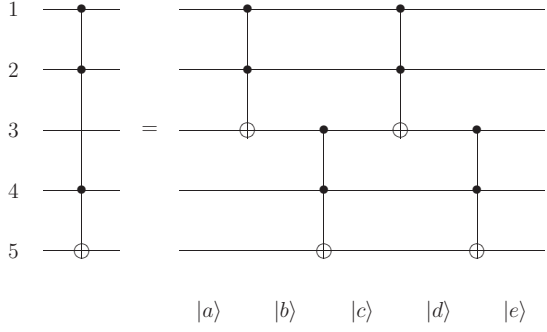}
\caption{\label{fig:c3not-to-toffoli} Decomposition of a $C^3$-NOT with one ancilla qubit
into four $C^2$-NOTs (Toffoli gates)
according to ref.~\cite{Barenco}. States $\ket{a}$ through $\ket{e}$ are explained in the text.
}
\end{center}
\end{figure}

To this end, we will generalise the basic principle of reducing a \cnnot to \cmnot gates with $m<n$ 
that can be demonstrated by decomposing a \cttnot into Toffoli gates according to scheme of 
Fig.~\ref{fig:c3not-to-toffoli} devised by Barenco {\em et al.} in \cite{Barenco}.
Starting with any of the $2^5$ computational basis states
$\ket{x_1,x_2,x_3,x_4,x_5}$ (where $x_k\in\{0,1\}$, $\oplus$ denotes addition $\mod 2$, and 
$x_k x_\ell$ being the usual scalar product) track the effect of the gate sequentially
from state $\ket{a}$ through state $\ket{e}$
\begin{enumerate}
\item[$\ket{a}$] $=\ket{x_1,x_2,x_3,x_4,x_5}$
\item[$\ket{b}$] $=\ket{x_1,x_2,x_3\oplus x_1x_2,x_4,x_5}$
\item[$\ket{c}$] 
$=\ket{x_1,x_2,x_3\oplus x_1x_2,x_4,x_5\oplus x_4(x_3\oplus x_1x_2)}$\\
$=\ket{x_1,x_2,x_3\oplus x_1x_2,x_4,x_5\oplus x_4x_3 \oplus x_1x_2x_4}$
\item[$\ket{d}$] 
$=\ket{x_1,x_2,x_3\oplus x_1x_2\oplus x_1x_2,x_4,x_5\oplus x_4x_3 \oplus x_1x_2x_4}$\\
$=\ket{x_1,x_2,x_3,x_4,x_5\oplus x_4x_3 \oplus x_1x_2x_4}$
\item[$\ket{e}$] 
$=\ket{x_1,x_2,x_3,x_4,x_5\oplus x_4x_3 \oplus x_4x_3 \oplus x_1x_2x_4}$\\
$=\ket{x_1,x_2,x_3,x_4,x_5 \oplus x_1x_2x_4}$
\end{enumerate}
to see the overall effect of the gate sequence is
a \cttnot thus proving the decomposition.

\begin{figure*}[Ht!]
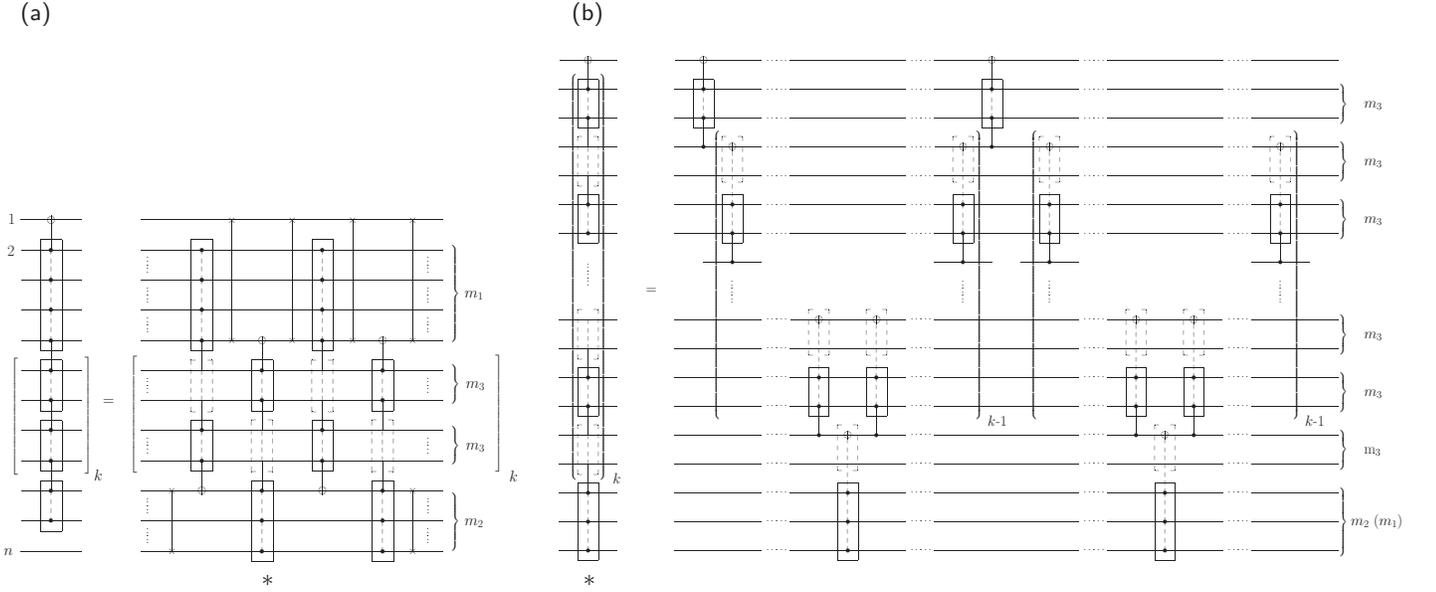

\begin{center}
{\sf \hspace{-102mm} (a) \hspace{67mm} (b)\hspace{0mm}}\\[3mm]
\mbox{
\hspace{-.4cm}
\raisebox{0mm}{\includegraphics[width=.385\textwidth]{BlockCnNOT.epsi}}
\hspace{.2cm}
\includegraphics[width=.63\textwidth]{BlockCnNOTb.epsi}
}\\
\end{center}
\vspace{-4mm}
{\sf \hspace{-71mm} $*$ \hspace{39mm} $*$}
\caption{\label{fig:dec_cnnot} Decomposition of a \cmtnot gate
on a linear coupling topology: (a) reduction of the number of control qubits to 
four intermediate gates with fewer control qubits and
(b) decomposition of the intermediate  multiply-controlled \gnot-gate
appearing in (a).
In an $n$"~qubit system, there is one target qubit,
one ancilla qubit and $n-2$ control qubits; so $m_1 + (m_2-1) + 2km_3 = n-2$
with $m_1, m_3 \geq 1$ and $m_2\geq 2$.
Read the brackets $[\cdot]_k$ in (a) as {\em to be expanded $k$ times}
and $(\cdot)_k$ in (b) as {\em expanded $k$"~fold}.
}
\end{figure*}
\begin{figure*}[Ht!]
{\sf \hspace{-45mm} (a) \hspace{81mm} (b)}
\begin{center}
\includegraphics[scale=.45]{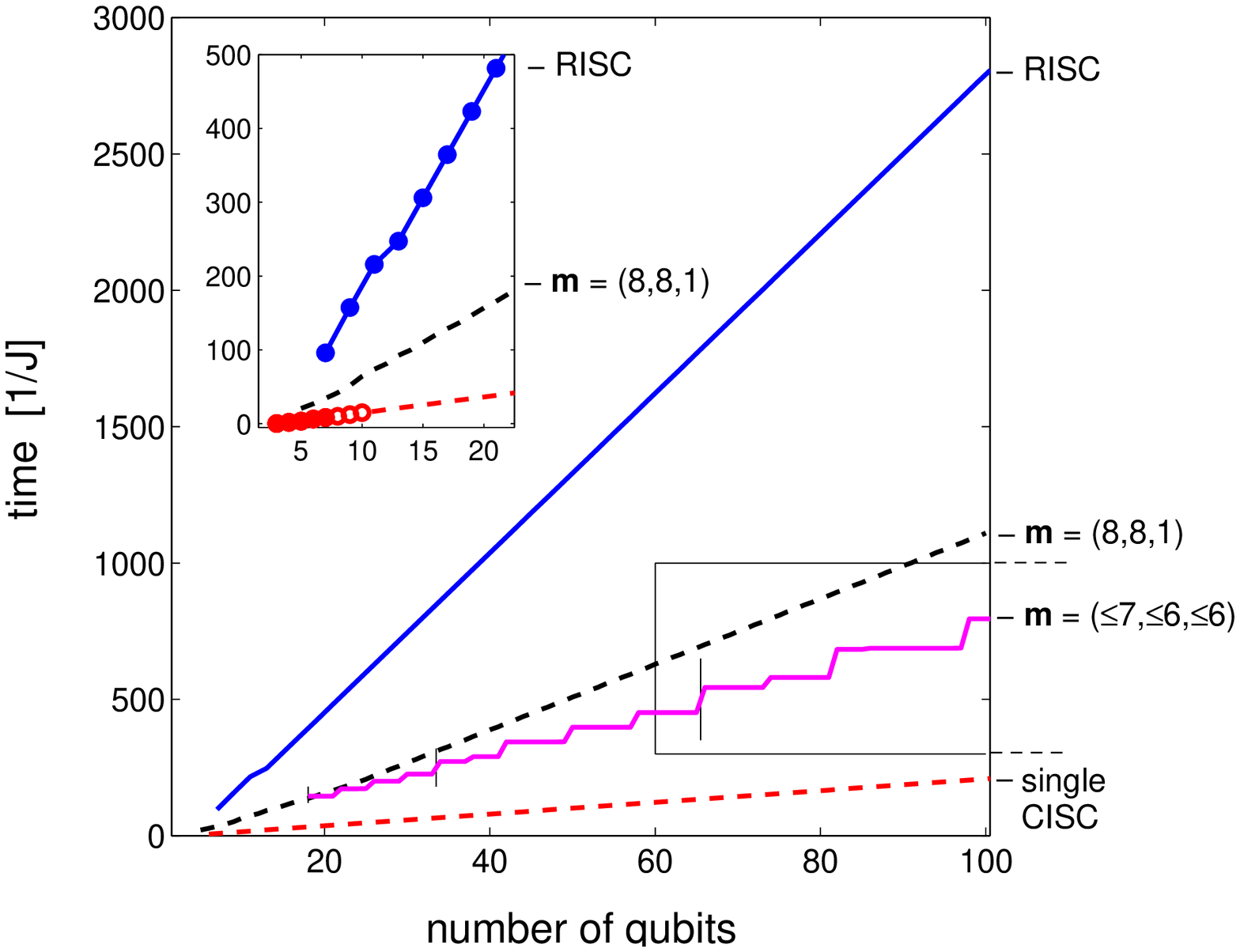}
\hspace{5mm}
\includegraphics[scale=.45]{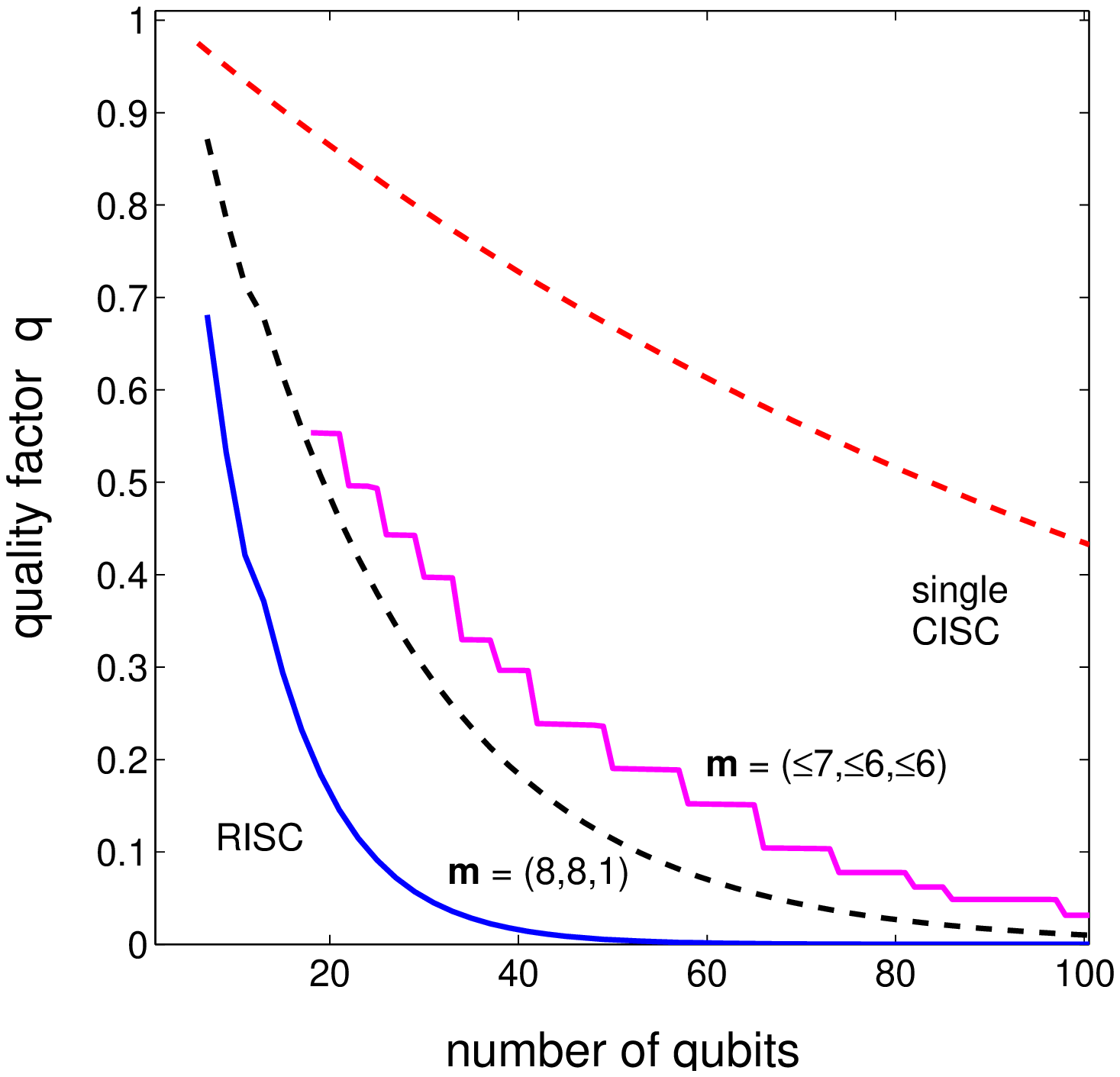}
\end{center}
\caption{\label{fig:CnNOTrec} (Colour online)
Comparing implementations of \cjnots {n-2} on a linear Ising spin chain using
\cnot and \swapj 2 modules for the \risc compilation or 
multi-qubit building blocks according to the \cisc assembler scheme of Fig.~\ref{fig:dec_cnnot}.
As a short-hand, the different numbers of control qubits are expressed by $\mathbf m := (m_1, m_2, m_3)$.
Using the expansion of Fig.~\ref{fig:dec_cnnot},
the \cisc results (black solid lines) are obtained for $k=4,5,6,\dots$ with $m_1 = 8$ and
$m_2 = 8$ (for odd $n$) or $m_2 = 7$ (for even $n$), while $m_3 = 1$ thus ensuring
$m_1 + (m_2 -1) + 2k m_3 = n-2$.
The red dotted line extrapolates again the direct \cisc results beyond 10 qubits.
In (a) deviations from straight lines occur, as the cases $k=1,2,3$ follow special concatenation
patterns (see text), while $k=4,5,6,\dots$ are generic.
The inset in (a) also shows results of a non-scalable recursive expansion that is confined
up to $19$ qubits (blue circles).
The step functions with periods indicated by tags represent a faster alternative explained 
in the next section, where the boxed part of trace (a) is blown up in Fig.~\ref{fig:CnNOTrec2}.
}
\end{figure*}

Fig.~\ref{fig:dec_cnnot} provides a generalisation 
of the scheme in Fig.~\ref{fig:c3not-to-toffoli}:
in the first place (a), the number of control qubits is reduced by introducing
$k$ blocks with $m_3$ qubits that are left invariant. The price for this
reduction is a four-fold occurence of the reduced building blocks.
In the second step (b), the reduced building blocks are expanded into
a sequence with two  central \cjnots {m_2}, two terminal \cjnots {(m_3 + 1)} 
and two lots of $2(k-1)$ times \cjnot {(m_3 + 1)} each.
For $k=4,5,6, \dots$ part (a) and (b) can be 
expanded in a general concatenated way thus entailing an overall duration of
\begin{equation}\label{eqn:tau_cnnot}
\begin{split}
\tau(&\cjnot{n-2})\Big|_{k\geq 4}\; =\; 
	 4 \tau(\cjnot{m_1}) \\[2mm]
        & + 4 \tau(\cjnot{m_2}) + \tau(\swapj{m_2}) \\[2mm]
        & + (13 k - 8) \tau(\cjnot{m_3+1}) \\[2mm]
	& + (1-\delta_{m_3, 1})\; (13 k + 3) \tau(\swapj{m_3})\;.\\[2mm]
\end{split}
\end{equation}
For completeness, note that the cases $k=3,2,1$ have to be treated separately,
since they only allow for less and less densly concatenated expansions (not shown). 
Their respective durations are
\begin{equation}\label{eqn:tau_cnnot1}
\begin{split}
        \tau(&\cjnot{n-2})\big|_{k=1}\; =\; 4 \tau(\cjnot{m_1})+2\tau(\swapj {m_1+1}) \\[.5mm]
        & + 4 \tau(\cjnot{m_2}) + 2\tau(\swapj {m_2}) \\[.5mm]
        & + 8 \tau(\cjnot{m_3+1}) + (1-\delta_{m_3, 1})\; 16 \tau(\swapj {m_3})\\[2mm]
\end{split}
\end{equation}
\begin{equation}\label{eqn:tau_cnnot2}
\begin{split}
        \tau(&\cjnot{n-2})\big|_{k=2}\; =\; 4 \tau(\cjnot{m_1})  \\[.5mm]
        & + 4 \tau(\cjnot{m_2}) + \tau(\swapj {m_2}) \\[.5mm]
        & + 24 \tau(\cjnot{m_3+1}) + (1-\delta_{m_3, 1})\; 32 \tau(\swapj {m_3})\\[2mm]
\end{split}
\end{equation}
\begin{equation}\label{eqn:tau_cnnot3}
\begin{split}
        \tau(&\cjnot{n-2})\big|_{k=3}\; =\; 4 \tau(\cjnot{m_1})  \\[.5mm]
        & + 4 \tau(\cjnot{m_2}) + \tau(\swapj {m_2}) \\[.5mm]
        & + 37 \tau(\cjnot{m_3+1}) + (1-\delta_{m_3, 1})\; 48 \tau(\swapj {m_3})\;.\\[2mm]
\end{split}
\end{equation}

However, the total number of gates only depends on $k=1,2,3,\dots$,
so that obtains as the overall quality
\begin{equation}\label{eqn:q_cnnot}
\begin{split}
q\big|_{k}\; = \; &(F_{C^{m_1}NOT})^4  (F_{\swapj {m_1+1}})^4\\[1mm]
        &\times (F_{C^{m_2}NOT})^4 (F_{\swapj {m_2}})^{2} \\[1mm]
        &\times (F_{C^{m_3+1}NOT})^{16 k - 8} \; (F_{\swapj {m_3}})^{(1-\delta_{m_3,1})\,16 k}\\[1mm]
        &\times e^{-\tau(C^{n-2}NOT)\big|_{k}/T_R}\;.
\end{split}
\end{equation}
Given the duration of the decomposition as in Eqn.~\ref{eqn:tau_cnnot}, it is easy to see that
implementing the $m_1$ control qubits comes with the lowest time weight (4) and without 
a time overhead of auxiliary gates. Implementing the $m_2$ control qubits, however, requires 
the same time weight (4), but entails the time for one auxiliary \swapj{m_2} gate.
In order to implement the $k\cdot m_3$ control qubits, in turn, a sizeable amount of auxiliary
\swaps are needed.

Therefore, whenever high fidelities can be reached (so that the quality is limited by 
relaxation not by fidelity),
a good strategy of combining the expansive decomposition in Fig.~\ref{fig:dec_cnnot}~a
with the recursive decomposition in part (b) is the following:
given $n-2$ control qubits and with the current limitation from direct \cisc compilation
being $m_j \leq 9$, choose $m_1$ to be the largest, $m_2$
to be the second largest and such that one obtains an even number for
$n - m_1 - m_2 -1 = 2k m_3$.

In the next step, a decision has to be made in order to minimise the contributions
in the last two lines of Eqn.~\ref{eqn:tau_cnnot}, whenever there are several integer solutions 
$k m_3 = k' m_3'$. So for integer $k\geq 4$, this amounts to the ordinary
minimisation task
\begin{equation}
\begin{split}
\minover{k}\; (13 k - 8)&\,\{(m_3+2)\Delta_C + a\} \\
	& \qquad + (1-\delta_{m_3,1})\; (13 k + 3)\,\{m_3 \Delta_S+ b\} \\[2mm]
\text{subject to:}\quad &km_3 = \text{const.} \equiv \frac{n - (m_1+m_2+1)}{2}
\end{split}
\end{equation}
Here we approximate the times for a \cjnot {m_3 + 1} by the linear expression 
$\tau(\cjnot {m_3+1}) = (m_3+2) \Delta_C + a$
and likewise for the \swapj {m_3} by 
$\tau(\swapj {m_3}) = m_3 \Delta_S + b$
with the values for the slopes $\Delta_C, \Delta_S$ and the offsets $a$ and $b$
being taken from the respective linear regression for extra\-polating 
$\Delta_\infty$ for direct \cisc compilation
($\Delta_C^{(\infty)}=2.15$ and $\Delta_S^{(\infty)}=0.69$ as well as $a=-4.48$ and $b=0.06$).
In the setting of these parameters,
Eqn.~\ref{eqn:tau_cnnot} implies it is timewise advantageous to choose
as the decomposition of the interior block in Fig.~\ref{fig:dec_cnnot}~b
the counter-intuitive option with a large number $k$ of small
block sizes $m_3$. This is because in the above parameter setting,
the duration takes its minimum on the margin circumventing the 
time overhead skipped by $(1-\delta_{m_3,1}) = 0$
thus giving high repetitions $k=\frac{n - (m_1+m_2+1)}{2}$ and 
smallest block sizes $m_3=1$ corresponding to Toffoli gates.
The speed-up is illustrated in Fig.~\ref{fig:CnNOTrec}: although it amounts to a factor
of $2.45$ compared to the standard \risc decomposition, the potential as extrapolated
from direct \cisc compilation up to nine qubits gives a lower bound for the speed-up
by $13.6$. 

\subsubsection*{Faster Alternatives of $\cjnot{n-2}$}

Since the potential of \cisc compiling \cnnot gates is largely
not yet exploited by the previous scheme, it is worth showing a 
faster scalable decomposition at the expense of being more elaborate.
To this end, we proceed
in two steps, first we show the general principle of an auxiliary
backbone gate, namely an indirect \cnot between qubit 1 and some distant 
qubit $\ell+1$ (which may be separated by $\ell$ intermediate qubits, e.g., in a linear
coupling topology). Second we implement
the resulting faster alternative into Fig.~\ref{fig:dec_cnnot}~b.
\begin{figure}[Ht!]
\begin{center}
\includegraphics[scale=.35]{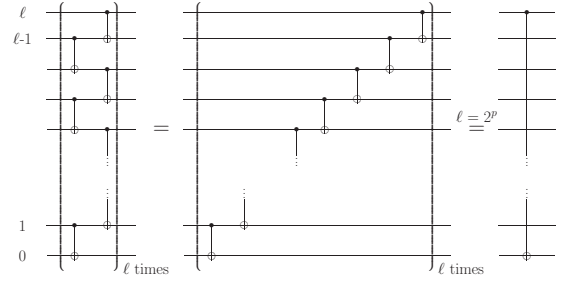}
\end{center}
\caption{\label{fig:CNOT_decomp} Principle of generating an indirect \cnot by restricting
$\ell=2^p$ as described in the text.
}
\end{figure}


Fig.~\ref{fig:CNOT_decomp} shows the principle identities: 
if $\ell$ is an integer power of two, the two identities hold for any $\ell=2^p$.
The second identity is easy to see since the ascending series of \cnot gates
can be represented by a Jordan matrix over the field of 
binary numbers $\Z_2^{\ell+1}:=\{0,1\}^{\ell+1}$ with the addition modulo $2$
so as to take the form
\begin{equation}
(J)_{\Z_2^{\ell+1}}=
\begin{pmatrix}
1 & 1 & 0&0& \cdots & 0& 0& 0\\
0 & 1 & 1&0& \cdots & 0& 0& 0\\
0 & 0 & 1&1& \cdots & 0& 0& 0\\
0 & 0 & 0&1& \cdots & 0& 0& 0\\
\vdots &  & && \ddots & & & \vdots\\
0 & 0 & 0&0& \cdots & 1& 1& 0\\
0 & 0 & 0&0& \cdots & 0& 1& 1\\
0 & 0 & 0&0& \cdots & 0& 0& 1
\end{pmatrix}^t \;.
\end{equation}
In terms of natural numbers, its $\ell^{\rm th}$ power reads
\begin{equation}
J^\ell =
\begin{pmatrix}
1 & \binom{\ell}{1} & \binom{\ell}{2}&\binom{\ell}{3}& \cdots & \binom{\ell}{\ell-2}& \binom{\ell}{\ell-1}& \binom{\ell}{\ell}\\[1mm]
0 & 1 & \binom{\ell}{1}&\binom{\ell}{2}& \cdots & \binom{\ell}{\ell-3}& \binom{\ell}{\ell-2}& \binom{\ell}{\ell-1}\\[1mm]
0 & 0 & 1&\binom{\ell}{1}& \cdots & \binom{\ell}{\ell-4}& \binom{\ell}{\ell-3}& \binom{\ell}{\ell-2}\\[1mm]
0 & 0 & 0&1& \cdots & \binom{\ell}{\ell-5}& \binom{\ell}{\ell-4}& \binom{\ell}{\ell-3}\\[1mm]
\vdots &  & && \ddots & & & \vdots\\[1mm]
0 & 0 & 0&0& \cdots & 1& \binom{\ell}{1}& \binom{\ell}{2}\\[1mm]
0 & 0 & 0&0& \cdots & 0& 1& \binom{\ell}{1}\\[1mm]
0 & 0 & 0&0& \cdots & 0& 0& 1
\end{pmatrix}^t \;.
\end{equation}
For $\ell=2^p$ with $p=1,2,3,4,\dots$ it gives
the desired indirect $\cjnot{1,\ell+1}$ as seen in the representation over $\Z_2^{\ell+1}$
\begin{equation}
(J^\ell)_{\Z_2^{\ell+1}} =
\begin{pmatrix}
1 & 0 & 0&0& \cdots & 0& 0& 1\\
0 & 1 & 0&0& \cdots & 0& 0& 0\\
0 & 0 & 1&0& \cdots & 0& 0& 0\\
0 & 0 & 0&1& \cdots & 0& 0& 0\\
\vdots &  & && \ddots & & & \vdots\\
0 & 0 & 0&0& \cdots & 1& 0& 0\\
0 & 0 & 0&0& \cdots & 0& 1& 0\\
0 & 0 & 0&0& \cdots & 0& 0& 1
\end{pmatrix}^t 
=(\cjnot{1,\ell+1})^t_{\Z_2^{\ell+1}}\; .
\end{equation}
This is because due to a theorem by Lucas \cite{LucasThmAll} only
for $\ell$ being an integer power of two, all the binomial coefficients
$\binom{\ell}{j}$ with $j=1,2,\dots,(\ell-1)$ are even,
while $\binom{\ell}{0}=\binom{\ell}{\ell}=1$. They are therefore
the only ones not to vanish in the representation over $\Z_2^{\ell+1}$.

\begin{figure}[Ht!]
{\sf \hspace{-6mm} (a) \hspace{21mm} (b)\hspace{0mm}}\\[3mm]
\begin{center}
\includegraphics[scale=.40]{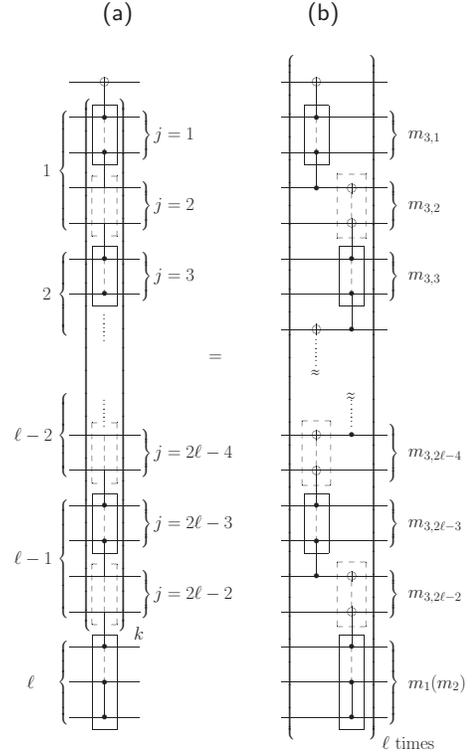}\hspace{3mm}
\end{center}
\caption{\label{fig:dec_cnnot2} (a) Alternative decomposition of the constituent module 
of Fig.~\ref{fig:dec_cnnot}~a. The qubits tagged by the numbers $1,2,\dots,\ell$ relate to
the neutral qubits of Fig.~\ref{fig:CNOT_decomp}.  (b) Each auxiliary building block involves a
solid box containing $m_{3,j+1}$ control qubits and a dashed box containing $m_{3,j}$
spacer qubits. It is termed 
$\cjnotk {(1+m_{3,j+1})} {({m_{3,j}})}$ and its realisation is shown in Fig.~\ref{fig:dec_cnnot2b}.
}
\end{figure}

\begin{figure}[Ht!]
\begin{center}
\includegraphics[scale=.33]{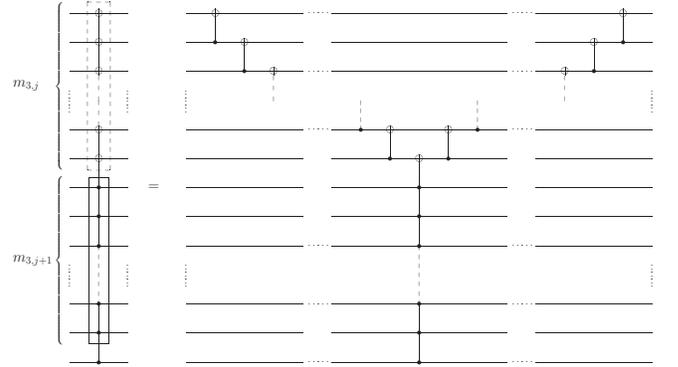}
\end{center}
\caption{\label{fig:dec_cnnot2b}  
Realisation of the auxiliary building block involving several {\sc not} actions.
}
\end{figure}

The principle backbone  summerised in the identities of Fig.~\ref{fig:CNOT_decomp}
may then be extended for more general purposes: (1) Without changing $\ell$ one may
insert further control qubits in the sense of replacing any \cnot by a Toffoli or
a higher $\cjnot m$. (2) Likewise, one may formally insert further target qubits to
be flipped so that the \qnot component is performed on more than one qubit.
These two extensions enable a faster alternative for decomposing the module of
Fig.~\ref{fig:dec_cnnot}~a than given in Fig.~\ref{fig:dec_cnnot}~b. This alternative
is shown in Fig.~\ref{fig:dec_cnnot2}~a.
Due to the backbone scheme of Fig.~\ref{fig:CNOT_decomp}, the only constraint
is that \/`1 plus the number of neutral blocks of size $m_{3,j}$
(represented by dashed boxes in Fig.~\ref{fig:dec_cnnot2}~a)\/´
equals $\ell=2^p$ with $p\in\N$\/'.
Changing the assembly of a $\cjnot {n-2}$ from the scheme of Fig.~\ref{fig:dec_cnnot} 
to the alternative of Fig.~\ref{fig:dec_cnnot2} follows by identifying
$k = \ell-1 = 2^p -1$, i.e. 
\begin{equation}\label{eqn:cnnot_var}
\begin{split}
n-2 &= m_1 + (m_2 -1) + 2k m_3 \\[2mm]
	&= m_1 + (m_2 -1) + \sum\limits_{j=1}^{2(2^p-1)}m_{3,j} \quad, 
\end{split}
\end{equation}
where we explicitly allow for individual block sizes $m_{3,j}$.
As shown in Fig.~\ref{fig:dec_cnnot2}~b, the decomposition 
of $m_{3,j+1}$ control qubits (solid boxes) and $m_{3,j}$ spacer qubits (dashed boxes)
leads to an auxiliary gate,
which we term $\cjnotk {(1+m_{3,j+1})} {({m_{3,j}})}$. It can be realised as in Fig.~\ref{fig:dec_cnnot2b}.
Note that the construction scheme of Fig.~\ref{fig:dec_cnnot}~a requires
to each solid box an equally sized dashed box.

In order to express the overall duration, we need the following
notation: let the array $\vec m_3:=(\vec m_3^1,\vec m_3^2,\vec m_3^3,\vec m_3^4)$ 
of total length $2\ell -2$
comprise the box sizes $m_{3,j}$ of Fig.~\ref{fig:dec_cnnot2}~b
grouped into the four subsets 
\begin{itemize}
\item[$\vec m_3^1:$] sizes of the $\tfrac{\ell}{2}$ solid boxes on the left,
\item[$\vec m_3^2:$] sizes of the $\tfrac{\ell-2}{2}$ solid boxes on the right,
\item[$\vec m_3^3:$] sizes of the $\tfrac{\ell}{2}$ dashed boxes on the left, and
\item[$\vec m_3^4:$] sizes of the $\tfrac{\ell-2}{2}$ dashed boxes on the right,
\end{itemize}
and let $\overline{m_{3}^s}$ be the largest entry in $\vec m_{3}^s, s=1,2,3,4$.
Then the duration of the decomposition
of a $\cjnot{n-2}$"~gate of Fig.~\ref{fig:dec_cnnot}~a according to 
Fig.~\ref{fig:dec_cnnot2} and \ref{fig:dec_cnnot2b}
reads
\begin{equation}\label{eqn:tau-cnnot}
\begin{split}	
\tau(\cjnot{n-2})=&\phantom{;+}2\cdot 2^p {\Big (} \tau(\cjnot {\overline{m_{3}^1}+1})+ \tau(\cjnot {\overline{m_{3}^2}+1})   \\
&\qquad \qquad \quad \; \; +2\maxn (\overline{m_{3}^3},\overline{m_{3}^4}) \cdot \tau(\text{\cnot}) {\Big )} \\
&+2\cdot 2^p {\Big (} \tau(\cjnot {\overline{m_{3}^3}+1})+ \tau(\cjnot {\overline{m_{3}^4}+1}) \\
&\qquad \qquad \quad \; \; +2\maxn(\overline{m_{3}^1},\overline{m_{3}^2}) \cdot \tau(\text{\cnot}) {\Big )} \\
& +  \tau_{\swap_{1,m_2}}\; .
\end{split}
\end{equation}

Obviously Eqn.~\ref{eqn:tau-cnnot} is symmetric under the exchanges
$\overline{m_{3}^1}\leftrightarrow \overline{m_{3}^2}$ and
$\overline{m_{3}^3}\leftrightarrow \overline{m_{3}^4}$, while the $\maxn$-functions 
break a full symmetry that would also require invariance under
$\overline{m_{3}^1}\leftrightarrow \overline{m_{3}^3}$ 
and $\overline{m_{3}^2}\leftrightarrow \overline{m_{3}^4}$.
Consequently, the broken symmetry imposes rules how to increase the
box sizes in a time saving way.
However, since the duration is limited by the largest box size in each time slot 
(left part and right part in Fig.~\ref{fig:dec_cnnot2}~b), 
one can fill the left and the right slots sequentially.

\begin{figure}[Ht!]
\begin{center}
\includegraphics[scale=.4]{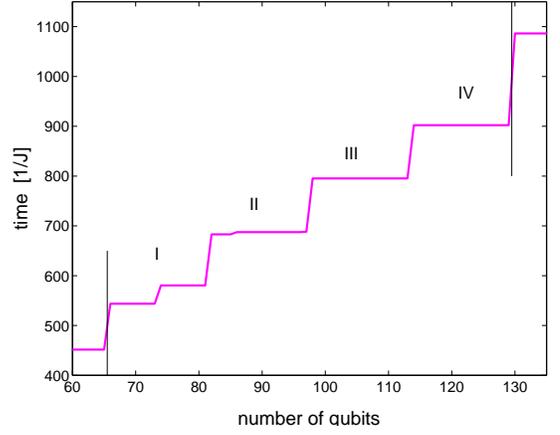}
\end{center}
\caption{\label{fig:CnNOTrec2}
(Colour online)
Detailed blow-up of the box in Fig.~\ref{fig:CnNOTrec}. The step function
is a periodic repetition of steps I-IV, where the length of the steps (in units of
numbers of qubits) doubles with every period as indicated by the tags in Fig.~\ref{fig:CnNOTrec}. 
The small jumps within steps I and II are explained in the text.
}
\end{figure}

Given $n-2$ control qubits, a time saving decomposition of a $\cjnot {n-2}$ results by
following the subsequent rules: calculate the auxiliary variables 
$p':=\lfloor \log_2(n-1)\rfloor -1$ and $r:=(n-1) - 2^{p'+1}$ to determine $p$ as 
\begin{enumerate}
\item $\mathbf{p=p'-1}$ for $p'\geq 2$ and $r\in [ 1, 2^{p'-2}]$ 
entailing $m_{3,j}\in\{2,3\}$ and the jump at half width
within step I (see Fig.~\ref{fig:CnNOTrec2});

\item $\mathbf{p=p'-2}$ for $p'\geq 3$ and $r\in [ 1+2^{p'-1}, 5\cdot 2^{p'-3}]$
entailing $m_{3,j}\in\{5,6\}$ and the minor jump at quarter width within step II (Fig.~\ref{fig:CnNOTrec2}); 

\item $\mathbf{p=p'}$ otherwise; then $m_{3,j}\in\{1,2\}$. 

\item NB: for $p=p'-3$ blocksizes would increase to $m_{3,j}\in\{8,9\}$ leading
to $\cjnot {9}$ and $\cjnot {10}$ building blocks, which are currently
out of reach.
\end{enumerate}
Then a time-saving decomposition obeys the final rule
\begin{enumerate}
\item[5.] Once $p$ and the box sizes $m_{3,j}\in\{b,b+1\}$ are fixed, arrange
the vector of grouped sizes
\begin{equation*}
\begin{split}
\vec m_3 &:= (\vec m_3^1,\vec m_3^2,\vec m_3^3,\vec m_3^4) \\
         & =(b+1,b+1,\dots,b+1,b,b,\dots,b,b)
\end{split}
\end{equation*}
with the entries in descending order.
\end{enumerate}
Clearly, the duration will not increase as long as 
all entries $(b+1)$ fall into $\vec m_3^1$, where one may choose
to start on top of Fig.~\ref{fig:dec_cnnot2}. A time-step
will occur as soon as the first $(b+1)$ falls into $\vec m_3^2$,
which neither affects $\maxn (\overline{m_{3}^1},\overline{m_{3}^2})$
nor $\maxn (\overline{m_{3}^3},\overline{m_{3}^4})$.
Analogous features hold for filling $\vec m_3^3$ and $\vec m_3^4$. They bring about the
periodic step function shown in Fig.~\ref{fig:CnNOTrec}.
Its details are given in Fig.~\ref{fig:CnNOTrec2}, where the
jump within step I is due to rule 1, while the
minor jump within step II has its roots in rule 2 above.

\medskip
\noindent
For illustration, consider the following three cases:

\begin{example}
In a system of $n=41$ qubits a $\cjnot {39}$ with one auxiliary qubit gives $p'=4$. 
By $r=8$ rule 3 applies and $p=4$. Eqn.~\ref{eqn:cnnot_var} and rule 5 then yield
$\vec m_3^1=(2)_8$ \footnote{The indices shall serve as a short-hand to denote, e.g.,
$(5)_2:=(5,5)\in\Z_2$ .}
and $\vec m_3^3=(1)_8$, while $\vec m_3^2=\vec m_3^4=(1)_7$
and $m_1=m_2=1$. 
Hence the time-saving decomposition involves just \cnot and Toffoli gates
(beyond the auxiliary \cnot and indirect \swap gates).
\end{example}

\begin{example}
Yet for $n=42$ qubits a $\cjnot {40}$ 
gives $p'=4$ and 
$r=9\in [1+2^{p'-2},5\cdot 2^{p'-4}]=[9,10]$,
so rule 2 applies and yields $p=2$.  By Eqn.~\ref{eqn:cnnot_var} and rule 5 one finds
$\vec m_3^1=(6,5)$ and $\vec m_3^3=(5,5)$ as well as 
$\vec m_3^2=(5)$ and $\vec m_3^4=(5)$ with $m_1=5$ and $m_2=5$. 
So the \cjnot {40} decomposes favourably via \cjnot 6 and \cjnot 7 gates.
\end{example}

\begin{example}
Finally, in a system of $n=137$ qubits a $\cjnot {135}$ gives $p'=6$ and $r=8\in [1,2^{p'-2}] = [1,16]$,
so rule 1 applies and $p=5$. Eqn.~\ref{eqn:cnnot_var} and rule 5 then give
$\vec m_3^1=\big((3)_8,(2)_8\big)$ and $\vec m_3^3=(2)_{16}$, $\vec m_3^2=\vec m_3^4=(2)_{15}$,
$m_1=m_2=2$. Therefore assembling a $\cjnot {135}$ refers to \cjnot 4 and \cjnot 3 gates.
\end{example}

%
%

Finally, the times of Eqn.~\ref{eqn:tau-cnnot} as well as the decomposition schemes
of Fig.~\ref{fig:dec_cnnot}~a, Fig.~\ref{fig:dec_cnnot2} and Fig.~\ref{fig:dec_cnnot2b} 
translate into the respective quality factors as
\begin{equation}
\begin{split}
&q(\cjnot{n-2})=
\left( F_{\cjnot {1+m_1}}\cdot F_{\cjnot {1+m_2}} \cdot F_{\cjnot {1+m_{3,1}}}\right)^{2^{p+1}} \\
&\times \prod_{j=2}^{\ell-1}\big\{\left( F_{\cjnot {1+m_{3,2j-1}}}\right)^{2^{p+1}}\cdot 
			\left( F_{\text{\cnot} } \right)^{2^{p+2} \cdot (m_{3,2j-2}-1)}\big\} \\
&\times \left( F_{\cjnot {1+m_{3,2\ell-2}}}\right)^{2^{p+1}} \\
&\times \prod_{j=2}^{\ell-1}\big\{\left( F_{\cjnot {1+m_{3,2\ell+2-2j}}}\right)^{2^{p+1}} \cdot
			\left( F_{\text{\cnot} } \right)^{2^{p+2} \cdot (m_{3,2\ell+1-2j}-1)}\big\} \\
& \times  \left( F_{\swap_{1,m_1+1}}\right)^4\phantom{;}\times \left( F_{\swap_{1,m_2}}\right)^2 \times e^{-\tau(\cjnot { n-2})/T_R}\;.
\end{split}
\end{equation}
The corresponding numerical quality results have already been
shown in Fig.~\ref{fig:CnNOTrec}~b as step function. They represent compilations with building blocks using
multiply controlled subblocks with up to $6$ and $7$ control qubits thus giving another significant
improvement over the assembly scheme described in the previous subsection. The results are also summarised
in Tab.~\ref{tab:cisc-results}.\\

\section{Implications for Multiply-Controlled General Unitary Operations}

The fast assembly schemes of
multiply controlled \gnot gates given in the previous subsection 
also allow for  faster realisations of 
multiply controlled general unitary gates than in the 
classic of Barenco, Bennett {\em et al.}, \cite{Barenco}.

\bigskip
\noindent
\begin{lemma}
Recall: every self-inverse 1-qubit special unitary $U_+=U_+^{-1}\in SU(2)$ is trivially $\pm\unity$, while every
self-inverse $U_-\in U(2)\setminus SU(2)$ is unitarily similar to $\sigma_x$.

\end{lemma}
{\bf Proof:}\quad  
To see the second assertion constructively, observe that
any self-adjoint $U_-\in U(2)\setminus SU(2)$ shows $\det U_-=-1$ and 
may thus be represented as a pure quaternion
$ U_- = x\cdot\sigma_x + y\cdot\sigma_y + z\cdot\sigma_z$
with $x^2 + y^2 +z^2 = 1$. 
Ensuring $|a|^2 + |b|^2 = 1$ in $V\in SU(N)$, it can be identified with 
\begin{equation}\label{eqn:sim-x}
 U_- = V\sigma_x V^\dagger = \begin{pmatrix} a & b \\ -b^* & a^* \end{pmatrix} 
	\begin{pmatrix} 0 & 1 \\ 1 & 0 \end{pmatrix}
	\begin{pmatrix} a^* & -b \\ b^* & a \end{pmatrix}
\end{equation}
via $x=\Re(a^2-b^2)$, $y = \Im(b^2-a^2)$, $z = 2\Re(ab^*)$.
$\hfill\blacksquare$
\bigskip

\begin{corollary}
Given a realisation of a $\cjnot{n-2}$ in time $\tau(\cjnot{n-2})$
on an $n$"~qubit system with one auxiliary and one target qubit.
Then the realisation of a multiply controlled general unitary $\cjU{n-2}$
takes the time
\begin{enumerate}
\item $\tau(\cjU{n-2}) \leq \tau(\cjnot{n-2}) +\tau(V) + \tau(V^\dagger)$,\\[2mm]
	 if $U\in U(2)\setminus SU(2)$ is self-inverse $U^2=\unity_2$
	 and $V\in SU(2)$ as in Eqn.~\ref{eqn:sim-x};

\item $\tau(\cjU{n-2}) \leq 2\cdot\tau(\cjnot{n-2}) + \tau(\text{$3$\, local\, gates})$,\\[2mm]
         if $U\in SU(2)$ and $U^2\neq\unity$;
\item Assertion (1) can be generalised to multiply-controlled $(q+1)$-qubit self-inverse unitaries of
	the form $U_- = V(\sigma_x\otimes\unity_{2^q}) V^\dagger$ with $V\in SU(2^{q+1})$.
\end{enumerate}
\end{corollary}
{\bf Proof:}\quad  

(1) The inequality is a direct consequence of applying Eqn.~\ref{eqn:sim-x} to the \gnot operation
on the target qubit.This is qubit $1$ in Fig.~\ref{fig:dec_cnnot}~a, which for later convenience
becomes qubit $0$ in Figs.~\ref{fig:CNOT_decomp} and \ref{fig:dec_cnnot2}. (Moreover, by reversing
Eqn.~\ref{eqn:sim-x} to $\sigma_x=V^\dagger U_- V$ and using it on qubit $0$ in Fig.~\ref{fig:CNOT_decomp},
one can absorb the time for at least one of the local gates $V$ by virtue of the decomposition of
Fig.~\ref{fig:dec_cnnot2}~b.)

(2) Direct consequence of Lemma~5.1 in Ref.~\cite{Barenco}.

(3) Obvious generalisation of Eqn.~\ref{eqn:sim-x} with $q$ further qubits
added---e.g., on top of qubit $1$ in Fig.~\ref{fig:dec_cnnot}~a.
$\hfill\blacksquare$

\medskip
\noindent
Similar generalisations hold for further special cases addressed in Ref.~\cite{Barenco}~Sec.~5.2.

%
%
\begin{table*}[Ht!]
\caption{\label{tab:cisc-results} 
Current Exploitation vs. Potential of Quantum \cisc Compilation (Limit of Fast Local Controls)
}
\begin{ruledtabular}
\begin{tabular}{llllll}
\\[-2mm]
gate & \multicolumn{1}{c}{\cisc potential:} & estimate %
	& \multicolumn{1}{l}{current status:} & exploitation & improvement\\[1mm]
& & \multicolumn{1}{l}{$\pi_\cisc=\Delta_2/\Delta_\infty$}  %
	&& \multicolumn{1}{l}{$\eta_m = \Delta_\infty/\Delta_m$} %
	& \multicolumn{1}{l}{$\xi_m = \Delta_2/\Delta_m$}\\[2mm]
\hline\\[-1mm]
\swapj n     & medium & 2.16  & fairly exhausted    & 0.86  \; [$m=8$] & $1.88$\\[1.5mm]
$n$-\qft     & medium & 2.27  & halfway exhausted      & 0.53 \;  [$\mathbf m = (5_{\qft},10_{\cpswap})$] & $1.20$\\[1mm]
\cjnot {n-2} & big    & 13.6  & not yet exhausted   & 0.18 \;  
		[$\mathbf m = (8,8,1)_{n\; {\rm odd}}$ or $(8,7,1)_{n\; {\rm even}}$] & $2.45$\\[1mm]
             &        &       &                     & 0.25--0.31 \; ${[\mathbf m = (\leq7, \leq 6,\leq 6)]}$ & 3.45--4.19\\[1mm]
\end{tabular}
\end{ruledtabular}
\end{table*}



\section{Conclusions and Outlook}
We have exploited the power of a cutting-edge high-performance parallel cluster
for quantum \cisc compilation. Thus the standard toolbox of universal quantum gate modules 
(\risc) has been extended by time-optimised effective multi-qubit gates (\cisc). We have
shown ways how these \cisc modules can be assembled in a scalable way in order to address
large-scale quantum computing on systems that are too large for direct \cisc compilation.
Although our optimal-control based \cisc-compilation routine exploits parallel matrix 
operations for clusters as well as fast matrix exponentials \cite{HLRB07}, increasing the
system size by one qubit roughly requires a factor of eight more \cpu time. Since 
direct \cisc compilation thus scales exponentially, scalable assembler schemes for 
multi-qubit \cisc modules are needed, and we have presented some elementary yet
important ones:

For indirect \swaps, the quantum Fourier transform, and multiply-controlled \gnot"~gates, the
\cisc decomposition is significantly faster than the standard \risc decomposition into local 
plus universal two-qubit gates. The current improvements range from $20 \%$ up to a speed-up
by nearly $300 \%$. However, as illustrated in Tab.~\ref{tab:cisc-results}, the potential of
\cisc compilation is by far not yet exhausted with the current schemes. ---
As a noteworthy side result, we have shown that gate errors in multi-qubit \cisc modules 
propagate more favourably than in \risc modules confined to two-qubit gates.

Assembling pre-compiled effective multi-qubit modules has further advantages beyond
saving gate time: a problem common to many implementations occurs as soon as smaller
decoherence-protected modules shall be embedded in larger effective systems. Usually
dissipative coupling to a new environment also introduces new sources of decay the original
module has not been optimised for. Therefore, practical handling is greatly facilitated, 
if the $m$-qubit modules with tailored optimisation under dissipation and decoherence
(see, e.g., \cite{PRL_decoh})
extend to larger units of relaxatively interacting qubits than the standard of $m=2$.
This advantage can readily be envisaged by a quantum-information processor, e.g., based on
trapped ions, where the \/`passive qubits\/' are stored with spatial separation from
the currently \/`active ones\/' in the processing unit 
(see, e.g., \cite{BW08,CBW06,MLW08} for overview and recent developments).

Moreover, controlling physical $m$-qubit modules will also allow for encoding 
logical qubits with specifically tailored optimisiation under more realistic 
relaxation models \cite{PRL_decoh} than in ideal \/`decoherence-free subspaces\/'.

This paves the way to another frontier of research: optimising the quantum assembler task on the extended
toolbox of quantum \cisc-modules with effective many-qubit interactions. Finally,
it is to be anticipated that methods developed in classical computer science
can also be put to good use for systematically optimising quantum assemblers.

\begin{acknowledgments}
This work builds upon parallel programming reported in Refs.~\cite{EP06,HLRB07}:
we wish to thank K.~Waldherr, T.~Gradl, and T.~Huckle for implementing further
inprovements beyond the above references.
The work has been supported in part by the integrated {\sc eu} project {\sc qap},
by the Bavarian programme of excellence {\sc qccc},
and by {\em Deutsche Forschungsgemeinschaft}, {\sc dfg},
within the incentive {\sc sfb}-631. Access to the high-performance parallel
cluster \hlrbii at {\em Leibniz Rechen\-zentrum} of the Bavarian Academy of
Science is gratefully acknowledged.
We wish to thank Matthias Christandl and
Robert Zeier for fruitful discussions.
\end{acknowledgments}

\bibliography{control21}

\begin{figure*}[Hb!]
\begin{center}
\section*{APPENDIX SECTION}
\subsection{Variant-II of Scalably Assembling a Quantum Fourier Transform}
\includegraphics[width=.95\textwidth]{BlockQFT.epsi}
\caption{\label{fig:qft-block}
For $k\geq 2$ and even $m$, a $(km)$"~qubit \qft can be assembled from $k$ times an $m$"~qubit \qft 
and $4 {k\choose 2}$ instances of $m$"~qubit modules \cpswapJM jm, 
where the index $j$ of different phase-rotation angles
takes the values $j=1,2,\dots, 2k -1$. 
The dashed boxes correspond to Fig.~\ref{fig:qft2-block} and show the induction $k\mapsto k+1$.
}
\end{center}
\end{figure*}
\begin{figure*}
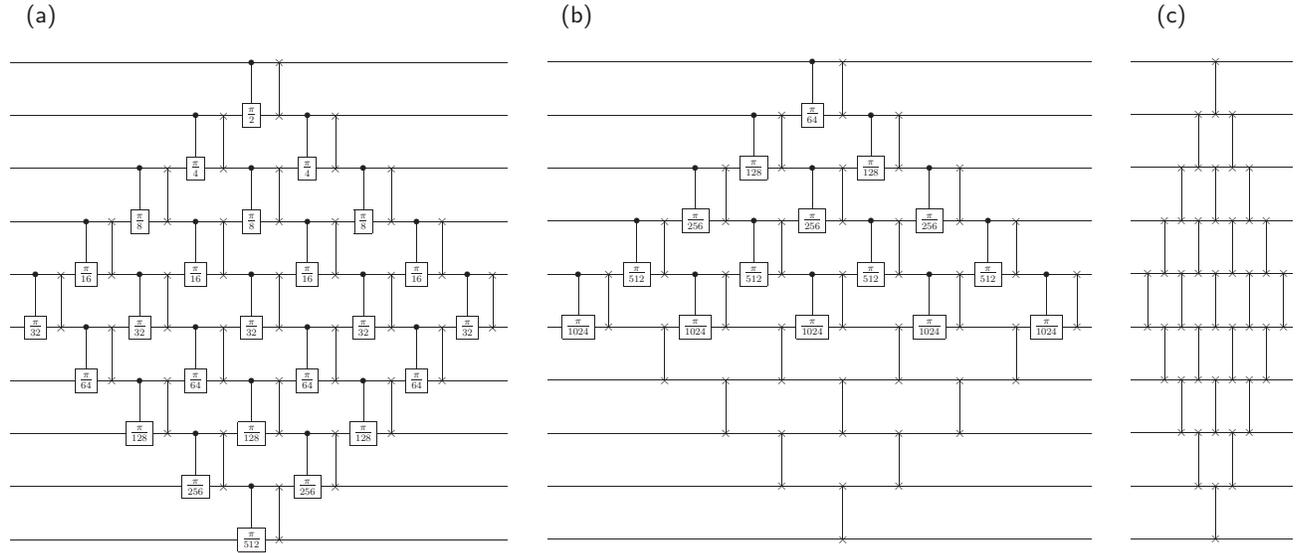

\subsection{Controlled-Phase-SWAP Modules for $k\cdot 10$-Qubit QFTs}
\mbox{\sf\hspace{7mm}(a)\hspace{67mm}(b)\hspace{75mm}(c)\hspace{52mm}}\\[3mm]
\begin{center}
\raisebox{-1.2mm}{
\includegraphics[scale=.463]{cPhaseSWAP1_10.epsi}\hspace{2mm}
}
\raisebox{.0mm}{
\includegraphics[scale=.463]{cPhaseSWAP2_10c.epsi}\hspace{2mm}
}
\raisebox{.07mm}{
\includegraphics[scale=.456]{cPhaseSWAP3_10.epsi}}

\end{center}
\caption{\label{fig:cp-swaps} Equivalent circuit representations of the three $10$-qubit
\cpswap modules needed for a $k\cdot 10$-qubit \qft, 
when rotation angles less than $\pi/2^{10}$ are omitted (as described in the text):
(a) \cpswapJM 1 {10} with no truncation so $F_{tr} = 1$, 
(b) \cpswapJM 2 {10} with $F_{tr} = 0.9999902$ , and (c) \cpswapJM j {10}, which covers all $j\geq 3$
with fidelity $F_{tr} \geq 0.9999991$.
These building blocks are compiled
directly as effective $10$-qubit modules thus reducing the time to less than $60 \%$ of
the duration required for the decomposition into $2$-qubit modules.}
\end{figure*}



\end{document}